\documentclass[prb,preprint,showpacs,preprintnumbers,amsmath,amssymb]{revtex4}
\usepackage{graphicx}% Include figure files
\usepackage{dcolumn}% Align table columns on decimal point
\usepackage{bm}% bold math
\usepackage{amsmath}
\usepackage[mathscr]{eucal}

\begin{document}
\preprint{preprint  ver: 4.3}

\title[Structure of $\alpha$-tetragonal boron]{Structure, non-stoichiometry, and geometrical frustration of $\alpha$-tetragonal boron}

\author{Naoki Uemura}
\author{Koun Shirai}
 \email{koun@sanken.osaka-u.ac.jp}
\affiliation{%
Nanoscience and Nanotechnology Center, Institute of Scientific and Industrial Research (ISIR), Osaka University, 8-1 Mihogaoka, Ibaraki, Osaka 567-0047, Japan
}%

\author{Hagen Eckert}
\affiliation{Institute for Materials Science and Max Bergmann, Center of Biomaterials, Dresden University of Technology, Germany}\author{Jens Kunstmann}
\affiliation{Theoretical Chemistry, Department of Chemistry and Food Chemistry, TU Dresden, 01062 Dresden, German}
\affiliation{Institute for Materials Science and Max Bergmann, Center of Biomaterials, Dresden University of Technology, Germany}
\date{\today}% It is always \today, today,

\begin{abstract} 
Recent discoveries of supposedly pure $\alpha$-tetragonal boron require to revisit its structure. The system is also interesting with respect to a new type of geometrical frustration in elemental crystals, which was found in $\beta$-rhombohedral boron. Based on density functional theory calculations, the present study has resolved the structural and thermodynamic characteristics of pure $\alpha$-tetragonal boron. 
Different from $\beta$-rhombohedral boron, the conditions for stable covalent bonding (a band gap and completely filled valence bands) are almost fulfilled at a composition B$_{52}$ with two $4c$ interstitial sites occupied. This indicates that the ground state of pure $\alpha$-tetragonal boron is stoichiometric. However, the covalent condition is not perfectly fulfilled because non-bonding in-gap states exist that cannot be eliminated. The half occupation of the $4c$ sites yields a macroscopic amount of residual entropy, which is as large as that of $\beta$-rhombohedral boron. Therefore, $\alpha$-tetragonal boron can be classified as an elemental crystal with geometrical frustration. Deviations from stoichiometry can occur only at finite temperatures. Thermodynamic considerations show that deviations $\delta$ from the stoichiometric composition (B$_{52+\delta}$) are small and  positive. For reported high-pressure syntheses conditions $\delta$ is predicted to be about 0.1 to 0.2. 
An important difference between pure and C- or N-containing $\alpha$-tetragonal boron is found in the occupation of interstitial sites: the pure form prefers to occupy the $4c$ sites, whereas in C- or N-containing forms a mixture of $2a$, $8h$, and $8i$ sites are occupied. The present article provides relations of site occupation, $\delta$ values, and lattice parameters, which enable us to identify pure $\alpha$-tetragonal and distinguish the pure form from other ones.

\end{abstract}

%Use showkeys class option if keyword
\maketitle

\section{Introduction}
\label{sec:intro}
\begin{figure}[htbp]
\centering
\includegraphics[bb={0 0 606 547},width=10 cm]{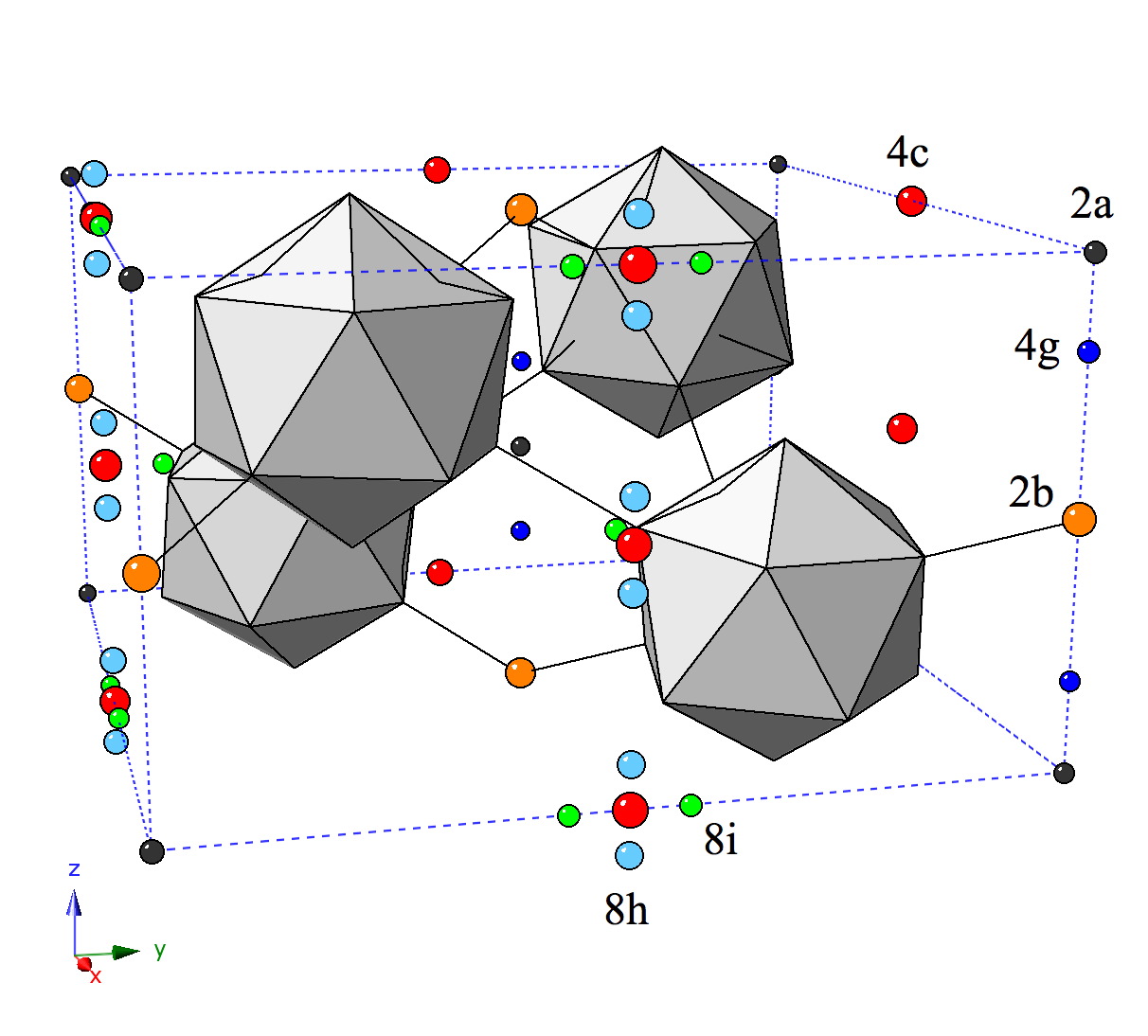}
\caption{The crystal structure of $\alpha$-tetragonal boron is composed of four icosahedral B$_{12}$ units and partially occupied interstitial sites that are indicated as colored balls. For a better visibility not all equivalent interstitial sites are shown. The sites are labeled according to the high symmetry space group ${\rm P4_{2}/nnm}$: $2a (000)$, $2b (00 \frac{1}{2})$, $4c (0 \frac{1}{2}0)$, $4g (00z)$, $8h (0\frac{1}{2}z)$, $8i (x 0 \frac{1}{2})$.} \label{fig:struct}
\end{figure}

The existence of $\alpha$-tetragonal ($\alpha$-T) boron has long been a controversial issue.
Historically, $\alpha$-T boron was first isolated in 1943 \cite{Laubengayer43}, and  tentatively identified as B$_{50}$, which is composed of four tetrahedrally coordinated B$_{12}$ icosahedra and two interstitial B atoms (see Fig.~\ref{fig:struct}).\cite{Hoard51} But soon this structural model was doubted by theorists. Longuet-Higgins and Roberts showed that the B$_{50}$ structure is unstable due to electron deficiency.\cite{LH55} Later, experimental groups showed that the actual crystals contained C or N atoms \cite{Amberger71,Ploog72,Will76} and the chemical compositions were approximately B$_{50}$N$_{2}$ or B$_{50}$C$_{2}$. This point of view was further supported by DFT calculations.\cite{Morrison92,Lee92a} Since then, it is almost generally accepted that pure $\alpha$-T form does not exist and that the structure is stabilized only by inclusion of foreign atoms. 

In the 2000's, nanostructures were repeatedly reported to have the structure of pure $\alpha$-T boron.\cite{Zhang02,Wang03,Xu04,Yang04,Kirihara05} Hayami and Otani explained the existence of pure $\alpha$-T boron in nanostructures by the presence of  low surface energies.\cite{Hayami08} Subsequently, they predicted the possibility of pure $\alpha$-T boron  bulk structures with the composition B$_{52}$.\cite{Hayami10}
After this theoretical work, the high-pressure syntheses of bulk $\alpha$-T boron were reported by several groups \cite{Ekimov11,Ekimov11a,Qin12,Kurakevych12,Kurakevych13,Solozhenko13}. Notwithstanding the prediction by Hayami and Otani, many different forms of tetragonal boron were discovered, including a known form of $\beta$-tetragonal boron \cite{Ma03}. The current situation occurs to be rather complicated and many questions are  raised: Does  pure $\alpha$-T boron truly exists?  And if it does, what stabilizes its structure and what was wrong with the previous theory? The purpose of this paper is to answer these questions. 

To say it shortly, the qualitative argument of Longuet-Higgins and Roberts is not wrong. The structure of B$_{50}$ is indeed not stable. However, there are ingenious ways to circumvent this instability.\cite{Shirai12} One way is a deviation from stoichiometry. A suitable deviation from stoichiometry can lead to complete valence band filling and thus stabilize the  structure of a covalent crystal. This explanation however raises a new question: Why was pure $\alpha$-T boron not found in bulk phases until recently? A simple answer is that there are other phases which are more stable at ambient conditions, {\it i.e.}, $\alpha$- and $\beta$-rhombohedral ($\alpha$-R and $\beta$-R) boron. However, the relative stability of these polymorphs can be changed at high pressures, which is a different way for stabilizing $\alpha$-T boron. To clarify the relative stability of boron phases, a detailed comparison is required, which will be addressed in another study. Here, we concentrate on $\alpha$-T boron, only and 
identify the best structure that pure $\alpha$-T boron can have, if it exists.

There are several theoretical studies on the structure of $\alpha$-T boron.\cite{Lee92a,Morrison92,Hayami10,Aydin11,Aydin12} However, these studies assumed stoichiometry. From our point of view \cite{Shirai12}, this assumption leads to an incorrect description of the metal/insulator behavior. 
The important role of non-stoichiometry has been recognized only recently and it will therefore be described in Sec.~\ref{sec:background}. 
Some words are also given there concerning  the terminologies related to defect states, because they are used differently in different fields of science.
This section also provides an overview and analysis of recent experiments, which is necessary because the current situation is very complicated. Through this analysis, we are able to delineate a clear approach for identifying pure $\alpha$-T boron.
After describing the calculation method (Sec.~\ref{sec:method}), the article studies  the electronic structures of individual interstitial sites (IS) in Sec.~\ref{sec:bondnature}.  By considering all the involved IS, the driving force for the deviation from stoichiometry is clarified in Sec.~\ref{sec:supercell}. Based on this analysis, a deviation is predicted for various synthesis conditions  (Sec.~\ref{sec:high-pressure}). An assessment on the characterization of recently discovered $\alpha$-T boron is attempted in Sec.~\ref{sec:lattice-para}. Behind the deviation from stoichiometry, we can say more about why $\alpha$-T boron is not satisfied with a simple structure B$_{52}$. Section \ref{sec:frustration} is devoted to discuss this topics from a view point of geometrical frustration. In the last section, we summarize our results.

\section{Theoretical background and experimental facts}
\label{sec:background}
% \section{Background}

For a long time solid-state theory of boron crystals had a fundamental problem. For many boron-rich crystals, band structure calculations showed metallic behavior, whereas in experiments all the crystals were found to be semiconductors. We call this problem the {\em metal/insulator} problem. Table \ref{tbl:stoichiometry} summarizes this discrepancy. This table is adapted from a paper of R. Schmechel and H. Werheit \cite{Schmechel99}, who first pointed out the importance of deviation from stoichiometry for the metal/insulator problem. The last entry of the table corresponds to $\alpha$-T boron, where for convenience the conclusion of this study is already given.
We can notice in this table that, whenever the metal/insulator problem occurs, deviation from stoichiometry is observed.
In boron-rich solids, the crystals are composed of a regular arrangement of icosahedra and various kinds of atoms at interstitial sites. Those interstitial sites are either fully occupied sites (FOS) or partially occupied sites (POS). The deviation from stoichiometry is a result of the presence of POS.

\begin{table}
  \caption{A comparison of hypothetical (theoretically considered) and real structures of boron-rich solids. 
For most systems theory was unable to correctly predict their semiconducting properties (second column). The experimentally determined electronic properties are given in the last column.
For the real structures, some ground-state properties are indicated, such as high- or low-symmetry (H-Sym./L-Sym.) and stoichiometric or non-stoichiometric compositions (St./NSt.). The experimentally determined electronic properties are given in the last column. $N_{\rm at}$ is the number of atoms per unit cell. Odd numbers of electrons $N_{\rm el}$ are indicated by italicized numbers. 
The last entry of this table contains our conclusion about the structure of $\alpha$-T boron, that is presented in this article.}
  \label{tbl:stoichiometry}
  \begin{tabular}{c | crrc | rcc}
    \hline \hline
    Crystal & \multicolumn{4}{c|}{Hypothetical structure} & \multicolumn{3}{c}{Real structure}  \\
                 &  Sym. & $N_{\rm at}$ & $N_{\rm el}$ & Prediction & $N_{\rm at}$ & Ground state & Elect. \\
    \hline \hline
    $\alpha$-rhomb.  &  $D_{\rm 3d}$ & 12 & 36 & semicon. & 12 & H-sym.  & semicon. \\
      &   &  &  &  &  & St. &  \\ \hline
    $\beta$-rhomb.  &  $D_{\rm 3d}$ & 105 & {\it 315} & metal & 106.5 & L-sym. & semicon. \\ 
      &   &  &  &  &  & NSt. &  \\ \hline
    ${\rm B_{13}C_{2}}$  &  $D_{\rm 3d}$ & 15 & {\it 47} & metal & 15 & L-sym. & semicon. \\ 
      &   &  &  &  & (on average) & NSt. &  \\ \hline
    $\alpha$-tetra.  &  $D_{\rm 4h}$ & 50 & 150 & metal & 52+$\delta$ & $\delta=0$ for $T=0$ & semicon. \\ 
       &   & 52 & 156 & semicon. &  & $\delta \neq 0$ for $T>0$ &  \\
    \hline \hline
  \end{tabular}
  % \textsuperscript{\emph{a}} comment;
\end{table}

It is well known that boron-based materials are characterized as {\it electron deficient} systems in the chemical literature.\cite{Pauling} Electron deficiency and the metal/insulator problem occur unrelated and therefore we had not be aware until recently that these two different matters are actually intimately related.
In the last decade, the effect of POS was extensively studied by DFT calculations on $\beta$-R boron.\cite{Masago06,Setten07,Widom08,Ogitsu09} It was found that the occupation of IS in boron-rich solids cannot be considered as defects, contrary to the usual notion in the literature about defect physics. The interstitial atoms are a part of the host crystals, in a sense that the energy is lower than that of the perfect crystals. The deviation from stoichiometry in $\beta$-R boron led Ogitsu and Widom to conceive a new concept of {\it geometrical frustration}.\cite{Ogitsu09,Ogitsu10,Widom08} Shirai and Uemura have finally established a relationship between the deviation from stoichiometry and the metal/insulator problem.

The mechanism behind the insulating property of boron crystals, which are otherwise predicted as metals, can be elucidated by the following way.\cite{Shirai12} (i) The system has an odd number of electrons ($N_{\rm el}$), which  is a consequence of electron deficiency. Then band theory unequivocally predicts it to be a metal. (ii) The system has strong covalency. This requires an even number of electrons. These two conditions (i) and (ii) are mutually incompatible. This competition yields a strong driving force to modify chemical bonds. Reconstruction of chemical bonds is usually difficult because of high energy barriers which exist between different configurations. By combining with the third condition (iii) of a large unit cell size (large number of atoms $N_{\rm at}$ per cell), the energy barrier can largely be reduced. The deviation from stoichiometry is an efficient solution for reconciling the electron deficiency and the covalency. Using this mechanism, difficult problems with the electronic properties 
of B$_{13}$C$_{2}$ were resolved.\cite{Shirai14,Ektrarawong15}

% \paragraph{$\alpha$-tetragonal boron case}
The above-mentioned mechanism should naturally work for $\alpha$-T boron, too. 
The initially assumed structure of $\alpha$-T boron is B$_{50}$, which is composed of four icosahedra and two $2b$-site B atoms (Fig.~\ref{fig:struct}). The $2b$ site is almost perfectly occupied (see Table \ref{tbl:alpha-tetra}). Therefore B$_{50}$ was believed to be stoichiometric. Later we will show, however, that this model should be revised.
Although the hypothetical structure B$_{50}$ has an even number of electrons and condition (i) does not hold, the valence-electron counting indicates that the covalent condition (ii) is not satisfied, because the system is short of ten electrons to completely fill up the valence band.
This is the reason why Longuet-Higgins and Roberts cast doubt on the existence of B$_{50}$.\cite{LH55} 
Their conclusion is essentially correct even in modern DFT calculations. See, for example, Fig.~3 of Ref.~23 where the top five bands of the valence band are unoccupied. 
 
Aside from the $2b$ site, many IS, which are partially occupied, have been reported (see Fig.~\ref{fig:struct}). The occupancies are listed in Table \ref{tbl:alpha-tetra}. As described in the Introduction, the $\alpha$-T boron crystals reported in the last century are now believed to contain C or N impurities and the crystals reported by Hoard \cite{Hoard58} may belong to this class, too. Hereafter we will call these crystals {\it traditional} $\alpha$-T boron. Such traditional $\alpha$-T boron always contains impurities such as C or N, and exhibits a deviation from stoichiometry. 

\begin{table}
  \caption{A comparison of different $\alpha$-tetragonal boron structures that were reported in the literature. The lattice parameters and occupancies of partially occupied sites vary greatly among the different samples. Hoard's and Will's forms are traditional $\alpha$-T boron, and the others  are recently discovered forms. The lattice parameters $a_0$ and $c_0$ are compared by evaluating the difference from Hoard's crystal in \%. For Qin's structure, the name ${\rm B_{50}}$ is only nominal and does not indicate the accurate composition.}
  \label{tbl:alpha-tetra}
  \begin{tabular}{c | r| r|r| r| r| r| r}
    \hline
    Author & Hoard \cite{Hoard58} & \multicolumn{2}{c|}{Will \cite{Will76} } & Hyodo \cite{HyodoTh} & Ekimov \cite{Ekimov11} & Qin \cite{Qin12} & Kurakevych \cite{Kurakevych12} \\
    \hline
    Formula &  ${\rm B_{50}}$ & ${\rm B_{50}C_{1.9}}$ & ${\rm B_{50}N_{1.8}}$ & ${\rm B_{52.2}}$ & ${\rm B_{51.5}}$ & ${\rm B_{50}}$  & various C conc. \\
      &   &  &  &  &  &  & including 0\% \\
    \hline 
    $a_{0}$ (\AA) &  8.750 & 8.753 & 8.634 & 8.808 & 9.0508 & 8.71 & $ 8.775 \sim 8.93$  \\ 
    \multicolumn{1}{r|}{(\%)} &  0 & $+0.03$ & $-1.3$ & $+0.63$ & $+2.8$ & $-0.46$ & $+0.28 \sim +2.1$ \\ 
    $c_{0}$ (\AA) &  5.060 & 5.093 & 5.128 & 5.047 & 5.1341 & 5.00 & $5.064 \sim 5.08$ \\ 
    \multicolumn{1}{r|}{(\%)} &  0 & $+0.65$ & $+1.3$ & $-0.27$ & $+1.5$ & $-0.01$ & $+0.08 \sim +0.40$ \\ 
    \hline
    Site & \multicolumn{7}{c}{Occupancy} \\ \hline
    $2a$  &  - & 12.8 & 1.4 & 11 & - &  &  \\ 
    $2b$  &  100 & (C) 90.8 & (N) 92.9 & 93 & 100 &  &  \\ 
    $4c$  &  - & - & - & 0 & 31 &  &  \\ 
    $4g$  &  - & - & - & - & 6 &  &  \\ 
    $4d$  &  - & - & - & 0 & - &  &  \\ 
    $8h$  &  - & 11.2 & 2.6 & 2 & - &  &  \\ 
    $8i$  &  - & 9.8 & 23 & 24 & - &  &  \\ 
    \hline \hline
  \end{tabular}
\end{table}

Recently, motivated by a quest for pure $\alpha$-T boron, Hayami and Otani showed that B$_{52}$, with two occupied $4c$ sites, is the lowest-energy structure, and they suggested that pure $\alpha$-T boron exists in bulk form.\cite{Hayami10} Hereafter, when speaking of non-stoichiometry, we refer to ${\rm B_{52}}$ as the stoichiometric composition, for a reason which will be clear below. The deviation from stoichiometry is expressed as ${\rm B_{52+\delta}}$, with a small fractional number $\delta$.
Incidentally, several articles reporting the synthesis of pure $\alpha$-T boron have been published.\cite{Ekimov11,Ekimov11a,Qin12,Kurakevych12,Kurakevych13,Solozhenko13}
These newly discovered $\alpha$-T forms should be seen as distinct from the traditional boron, because in most cases they were obtained by solid-state phase transformation from $\beta$-R boron. Thus they are very likely to be truly pure boron crystals.
Unfortunately, the chemical compositions in the new $\alpha$-T borons are not well analyzed. For example, Qin's $\alpha$-T boron was designated as ${\rm B_{50}}$.\cite{Qin12} However, this is only because they could not measure the chemical composition accurately. Owing to the lack of reliable characterization, it is not even certain  if the reported $\alpha$-T borons are all the same. In this paper, we shall refer to the reported $\alpha$-T borons by the name of the first author, for example, Ekimov's $\alpha$-T boron. Kurakevych's $\alpha$-T borons are actually a series of crystals containing a C coagent ranging from 0 to 5 at.\%.\cite{Kurakevych12} 

Hyodo analyzed Kirihara's samples, reporting a slight non-stoichiometry ($\delta$=+0.2).\cite{HyodoTh} However, Kirihara's $\alpha$-T boron comes from nanostructures\cite{Kirihara05} and therefore his analysis will not be discussed on the same ground as the results for bulk crystals.
Only Ekimov {\it et al} clarified the chemical composition of their bulk $\alpha$-T boron, the value being 51.5 ($\delta$=-0.5).\cite{Ekimov11,Ekimov11a} As shown in Table \ref{tbl:alpha-tetra}, Ekimov's $\alpha$-T boron has the largest lattice parameter $a_{0}$ ever obtained. Our calculations show that the lattice parameters shrink when C or N impurities occupy the $2b$ site. This was also reported experimentally.\cite{Ekimov13} From this, Ekimov {\it et al} claimed that their crystals are pure $\alpha$-T boron. Because of their synthesis method (pyrolysis of decaborane ${\rm B_{10}H_{14}}$) the possibility of hydrogen incorporation is not excluded. Their chemical analysis, however, did not indicate H content as high as it influences the lattice parameter.

On the basis of the above analysis, we can summarize the current state of $\alpha$-T boron research as follows: If the recently synthesized  $\alpha$-T borons are pure ones, there is little doubt that they are non-stoichiometric; in contrast theory predicts stoichiometric B$_{52}$ to be the lowest-energy structure. The question is thus, if the deviation from stoichiometry in real crystals is an intrinsic property (lowest-energy state) or not. For $\beta$-R boron\cite{Setten07,Widom08,Ogitsu09} and ${\rm B_{13}C_{2}}$ \cite{Shirai14}, the deviation is an intrinsic property. 
If this is not the case for $\alpha$-T boron, the deviation must be an extrinsic property, caused by entropic effects at high temperatures. Then the degree of deviation would depend on the preparation conditions. Accordingly, one should characterize the real crystals with respect to the preparation conditions.

\section{Computational Methods}
\label{sec:method}

The electronic structures of $\alpha$-T boron were studied by density functional theory using a pseudopotential method and the Osaka2k code.\cite{Osaka2k} It uses the LDA parameterization by Perdew and Zunger \cite{PZ81}, the Perdew-Burke-Ernzerhof form of the generalized gradient approximation (GGA) \cite{PBE96} and Troullier-Martins pseudopotentials \cite{TM91} with the fully separable Kleinman-Bylander form \cite{KB82}.
In all the cases, the kinetic cutoff energy was 40 Ry. Various $k$-point sampling methods were used. For calculations using the primitive unit cell, a $4 \times 4 \times 4$ grid was used and for supercell calculations $\Gamma$-point only sampling. The convergence was well tested in our previous studies.\cite{Masago06,Dekura11,Shirai14} 
The formation energy $\Delta E_{f}$ is defined as the difference of the total energy with respect to a reference state, which was one form of $\alpha$-T boron B$_{m}$ (the stoichiometric composition, $m$=52, in most cases). 
Then, $\Delta E_{f}$ of a composition B$_{m+n}$ is obtained by,
\begin{equation}
  \Delta E_{f}[{\text B}_{m+n}] = E[{\text B}_{m+n}] - \frac{m+n}{m}E[{\text B}_{m}],
  \label{eq:formE}
\end{equation}
where $E[{\text B}_{m}]$ is the total energy of B$_{m}$. With this definition of $\Delta E_{f}$, a negative value implies a more stable structure.
A similar definition is used to indicate the change of the specific volume $\Delta V$, where $E$ is substituted by $V$ in Eq.~(\ref{eq:formE}) and $\Delta V$ is evaluated in units of eV/GPa. Structural optimizations were performed with respect to atomic positions and cell parameters and no constraints on the crystal symmetry were imposed, except the tetragonal symmetry for the lattice parameters. Optimizing the cell parameters is important in view of the study of high-pressure phases. LDA was used only for supercell calculations, because of its computational efficiency, otherwise GGA was used. Of course, in this case, comparisons were made only between cells with the same size.

Comparisons between impurity-containing $\alpha$-T boron (impurity $X$=C or N) were also performed when necessary. In this case, compositions ${\rm B_{50}X_{2}}$ were assumed, where $X$-atoms were place on the $2b$ sites and the interstitial B atoms were placed either at the $8h$ or the $8i$ sites.

\section{Results and Discussion}
% \subsection{Bonding nature of interstitial sites}
\subsection{The structure of $\alpha$-tetragonal boron}
\label{sec:bondnature}

\subsubsection{Formation energy of interstitial sites}
\label{sec:formaton-energy}
The formation energies of various IS were extensively studied by Hayami and Otani.\cite{Hayami10} In this work we want to study the roles of IS in more detail. 
We constructed a B$_{50}$ structure, where the $2b$ sites are fully occupied, and then placed B atoms at various IS, one by one. The results for $\Delta E_{f}$ are shown in Table \ref{tbl:HfofB50}. 
However, care is needed for the interpretation of these values because usually an energy barrier is to be expected between two local energy minimum sites. Between the $4c$ and $8i$ sites, Hayami and Otani reported a small energy barrier, less than 10 meV. \cite{Hayami10} The values for $\Delta E_{f}$ in Table \ref{tbl:HfofB50} were obtained by performing steepest-descent minimization with typically seven or eight iterations, which usually is enough to locate local minimum configurations.
However, started from an $8i$ site, we found that continuing the iteration by more than 30 times finally brought the $8i$-site atom to the $4c$ site (see Supplemental Material). In this sense, $8i$ or $8h$ are metastable sites. However, to facilitate comparison with other calculations and experiments, we will still consider these sites as distinct, below.

The configuration B$_{52}$ with two $4c$ sites in the out of plane configuration (two $4c$-site atoms located in different $4c$ planes and  not being neighbors) is the lowest-energy configuration. Hereafter, we designate this configuration as B$_{50}$ + 2B$_{4c}$. As mentioned above, all formation energies are defined with respect to this configuration.

Let us now consider only the blocks with $N_{\rm at}=51$ or less in Table \ref{tbl:HfofB50}. We see that $\Delta E_{f}$ of many IS are lower than that of B$_{50}$ (3.14 eV).
This fact already indicates that these interstitial B atoms are not defects but are a part of the host structure. The order of the stability of these IS is $4c$ (-1.45), $8i$ (-1.35), and $8h$ (-1.28) (the values in parenthesis are differences in $\Delta E_{f}$ with respect to  B$_{50}$). The $2b$ site is always a stable site, as seen from its perfect occupancy, i.e., removing a $2b$-site atom from B$_{50}$ requires an energy of +1.45 eV.
The sites $2a$ (0.66) and $4g$ (0.98) are not stable. 

\begin{table}
  \caption{The formation energy $\Delta E_{f}$ of interstitial sites in $\alpha$-T boron within the primitive unit cell. The results indicate that B$_{52}$, with two 4c sites occupied, is the lowest-energy structure. The reference for  $\Delta E_{f}$ and the change in volume $\Delta V$ is  B$_{52}$ with two $4c$-site atoms in the out-of-plane configuration. In the second column '-2b' indicates a removal of a 2b atom. $2\times 4c$ should be read as two atoms at $4c$ sites. All the data are obtained by GGA, while LDA results are added for $\Delta E_{f}$ in parentheses. }
  \label{tbl:HfofB50}
  \begin{tabular}{c |c rrr}
    \hline \hline
    $N_{\rm at}$ & Sites & \multicolumn{2}{c}{$\Delta E_{f}$} & $\Delta V$  \\
      &   & \multicolumn{2}{c}{eV} & (eV/GPa) \\
    \hline
49 & $-2b$ & 4.59  & (4.76) & 0.068 \\ \hline
50 &   & 3.14 & (3.40) & 0.052 \\ \hline
51 & $4c$ & 1.69 & (1.66) & 0.034 \\
     &	$8h$ & 1.72 & (1.69) & -0.047 \\
     & $8i$ &	1.80	& (1.81) & 0.035 \\
     & $4g$ & 4.12	& (4.13) & 0.036 \\ 
     & $2a$ & 3.80 & (3.52) & 0.049 \\ \hline
52 & $2\times 4c$ (out-of-plane) & 0.0 & (0.0) & 0.0 \\
     & $2\times 4c$ (in-plane) & 0.19 & (0.15) & -0.006 \\
     & $2\times 4c$ (nearest) & 1.80 & (1.82) & -0.006 \\
     & $4c+8i$  & 0.19 & (0.13) & -0.005 \\
     & $4c+8h$  & 1.46 & (1.49) & 0.021 \\ \hline
53 & $3\times 4c$  & 1.31 & (1.11) & -0.028 \\
     &  $2\times 4c+8i$ & 2.68 & (2.65) & -0.036 \\
     &  $2\times 4c+8i'$ & 2.67 & (2.65) & -0.037 \\
     &  $2\times 4c+8h$ & 1.33 & (1.10) & -0.024 \\ \hline
54 &  $4\times 4c$ & 2.40 & (2.00) & -0.069 \\
    \hline \hline
  \end{tabular}
\end{table}

\begin{figure}[htbp]
\centering
\includegraphics[bb={0 0 682 674},width=4cm]{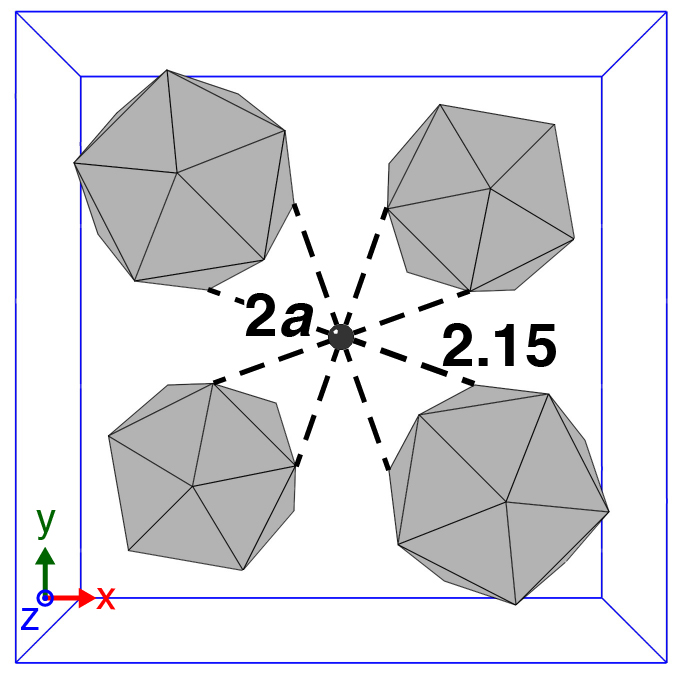}
\includegraphics[bb={0 0 682 674},width=4cm]{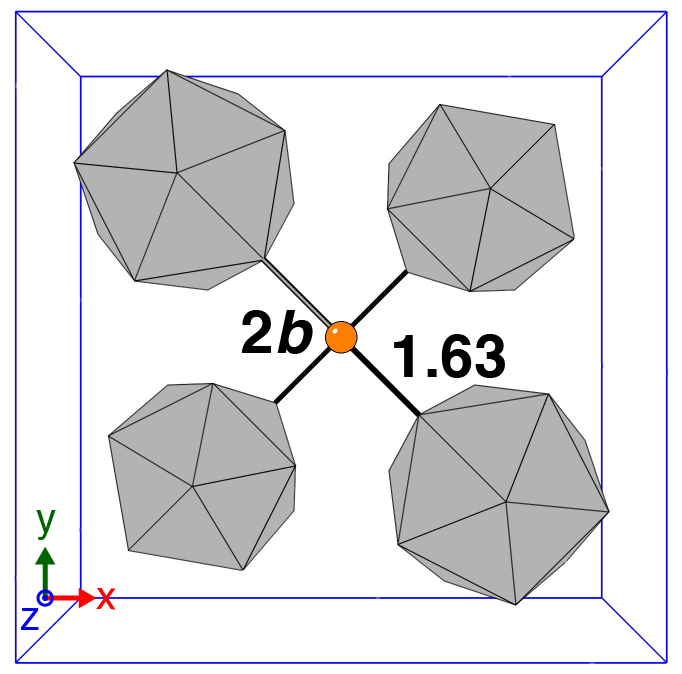}
\includegraphics[bb={0 0 682 674},width=4cm]{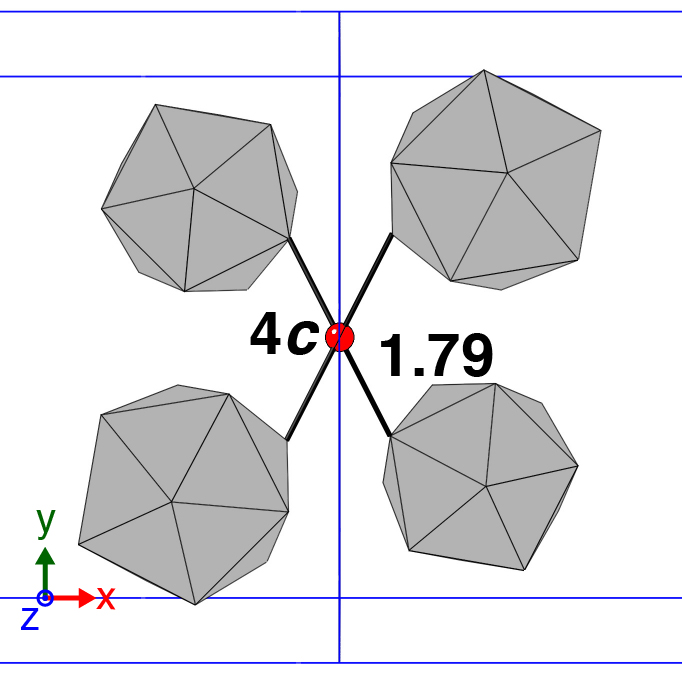}\\
\includegraphics[bb={0 0 682 674},width=4cm]{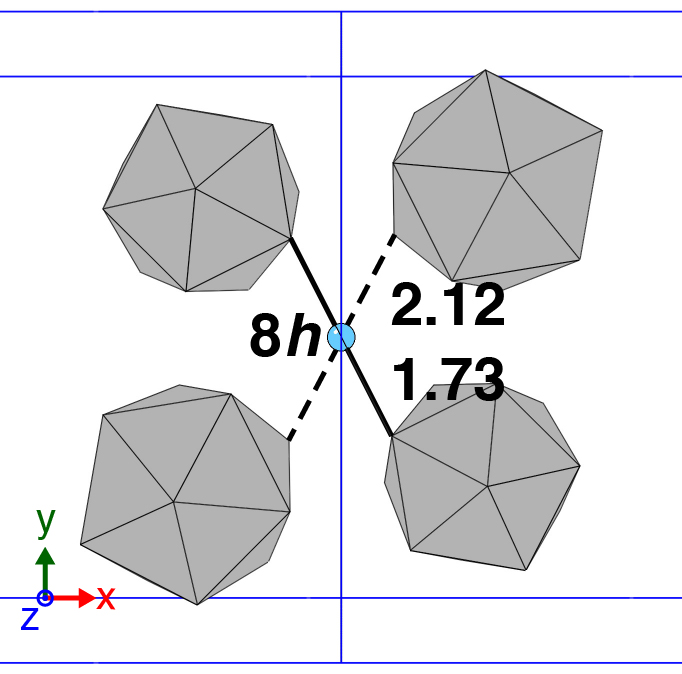}
\includegraphics[bb={0 0 682 674},width=4cm]{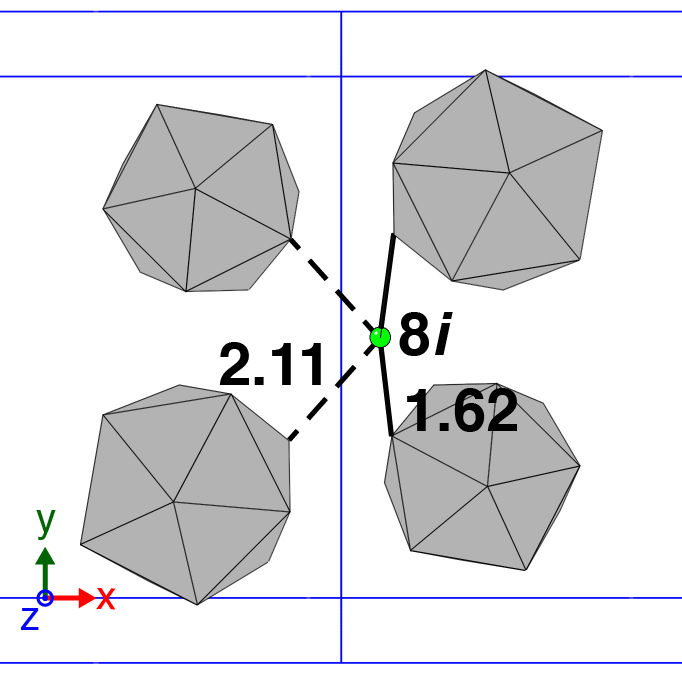}
\includegraphics[bb={0 0 682 674},width=4.3cm]{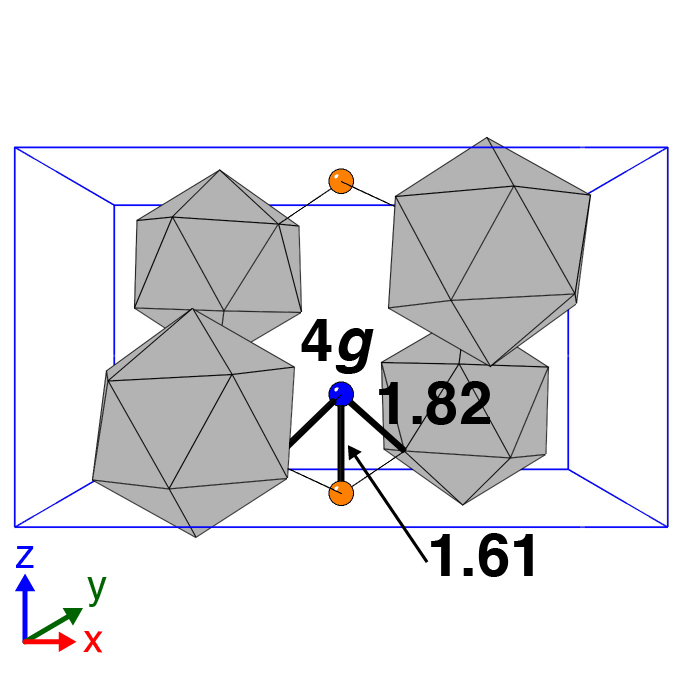}
%\hspace{1cm}
\caption{The partially occupied sites of $\alpha$-tetragonal boron and their local bonding environments. Most of the sites have tetragonal coordination. The experimental data obtained from Will's crystals \cite{Will76} are used. Bond lengths are given in \AA. The color code corresponds to Fig.~\ref{fig:struct}.}  \label{fig:POS}
\end{figure}

The energy ordering in $\Delta E_{f}$ can be understood by inspecting the bonding environment of these IS as shown in Fig.~\ref{fig:POS}. 
As seen, the $2b$ site is most perfectly tetrahedrally coordinated, which is the best configuration for a $sp^{3}$-type covalent bond. The short bond length of 1.63\ \AA \ also supports its strong covalent character. Hence, it is reasonable to obtain the perfect occupancy for this site. The $2a$ site also has perfect tetrahedral coordination, too. But, the nearest neighbor distance of 2.15\ \AA \ is too long. Furthermore, none of the atoms of the neighboring B$_{12}$ icosahedron are oriented in the direction linking the $2a$ site to B$_{12}$. Therefore, the bonding of $2a$-site atoms is weak. The site $4c$ is also close to tetrahedral coordination with a bond length of 1.79\ \AA, however the bond angles are largely distorted in the $ab$ plane. Although the $8i$ and $8h$ sites have short bond lengths (1.62 and 1.73\ \AA, respectively), there are only two of these short bonds, and the overall bonding is weaker than of $4c$-site atoms. 

To summarize, the IS of $\alpha$-T boron are not defect states. Their role is not just to fill space between the icosahedra, but it is a more active one, i.e., the enhancement of the host structure by forming tetrahedral bonds. The closer the sites are to the perfect tetrahedral coordination, the stronger is the bonding.

\subsubsection{Comparison with C- or N-containing $\alpha$-tetragonal boron}
\label{sec:impurity-tetra}
Although the positions of $8h$ sites $(0,1/2,u)$ and $8i$ sites $(u,0,1/2)$ are distinct  from $4c$ sites $(0,1/2,0)$ by symmetry, they can be continuously connected to the $4c$ site by varying $u$. However, the bonding environments of $8h$ and $8i$ are very different from that of $4c$. The coordination of the former is two-fold, while the latter is four-fold. 
For pure $\alpha$-T boron, $\Delta E_{f}$ of the $8i$ and $8h$ sites is bigger than that of the $4c$ sites by 0.03 and 0.11 eV, respectively.
On the other hand, for C- or N-containing $\alpha$-T boron, the energy difference between these sites almost vanishes (less than 0.001 eV or even negative). Only the energy of the $8h$ site of the N-containing form is higher by 0.008 eV.
The present calculations imply that the $4c$ site is the major interstitial for pure $\alpha$-T boron, whereas for C/N-containing $\alpha$-T boron comparable amounts of $8i$-site atoms are present and the occupation of the $8h$ site is the smallest.

This different order of $\Delta E_{f}$ for $4c$ and $8i$ sites between pure and C/N-containing $\alpha$-T boron is consistent with the experiment in Table \ref{tbl:alpha-tetra}. If the recently discovered $\alpha$-T forms are regarded as pure: $8h$ and $8i$ sites are observed in impurity-containing forms and $4c$ sites are observed in the pure form.
In our calculations, the bond lengths of C or N at the $2b$-site to their neighbors are short, {\it i.e.}, 1.57 and 1.59 \AA, respectively. The corresponding value for pure B$_{50}+ 2$B$_{4c}$ is 1.70\ \AA. In general, in icosahedron-based boron crystals, {\it inter}icosahedral bonds are the shortest ones but the value is at most 1.69 \AA. The experimental value of 1.63 \AA\ for traditional $\alpha$-T boron (see Fig.~\ref{fig:POS}) is too short for B-B bonds, so that the measured value may be evidence for C or N at the $2b$ site.

A clearer contrast between pure and C/N-containing forms is found in the occupation of the $2a$ site. 
$\Delta E_{f}$ of the pure form is large (3.80 eV), while that of the latter one is small (0.3 eV), so that for the pure form there are practically no $2a$-site atoms. Again, if recently discovered $\alpha$-T forms are pure ones, the present results for the site occupancies are consistent with the experiment, as seen in Table \ref{tbl:alpha-tetra}.

The different site occupations, {\it i.e.}, $4c$ in the pure form and  $8h$, $8i$, and $2a$ in the impurity-containing forms, provide a useful guideline for identifying impurities.  From the site occupations, we can say that at least Ekimov's $\alpha$-T boron is not a traditional $\alpha$-T form. To identify it as pure $\alpha$-T boron, however, we may need further evidence, because here the term 'impurity-containing form' is used in a restricted sense,, referring to C- or N-impurities, only. We cannot exclude the existence of other impurities. Further assessments of Ekimov's $\alpha$-T boron are made below.

\subsubsection{Valence filling}
\label{sec:valenc-filling}

Let us now consider to vary the composition $n$ in B$_{50+n}$ (hereafter, B$_{50+n}$ is also designated as B$_{50}+ n$B$_{s}$ to indicate $n$ atoms at site $s$). This is seen by inspecting the whole range of $N_{\rm at}$ in Table \ref{tbl:HfofB50}.
Hayami and Otani studied IS of $\alpha$-T boron over a compositional range $n=0-4$, concluding that B$_{50}+2$B$_{4c}$ is the most stable structure.\cite{Hayami10} The present calculations confirm their conclusion. Among various combinations of two $4c$-site atoms, the out-of-plane configuration has the lowest energy (For details of the structures, see Supplemental Material). A configuration in which two $4c$-site atoms occupy neighboring sites is highest in energy. 
There is a general tendency in boron-rich crystals that atoms of the same IS avoid each another. The present result for $4c$ sites is in accordance with this tendency.

\begin{figure}[htbp]
\centering
\includegraphics[width=8 cm]{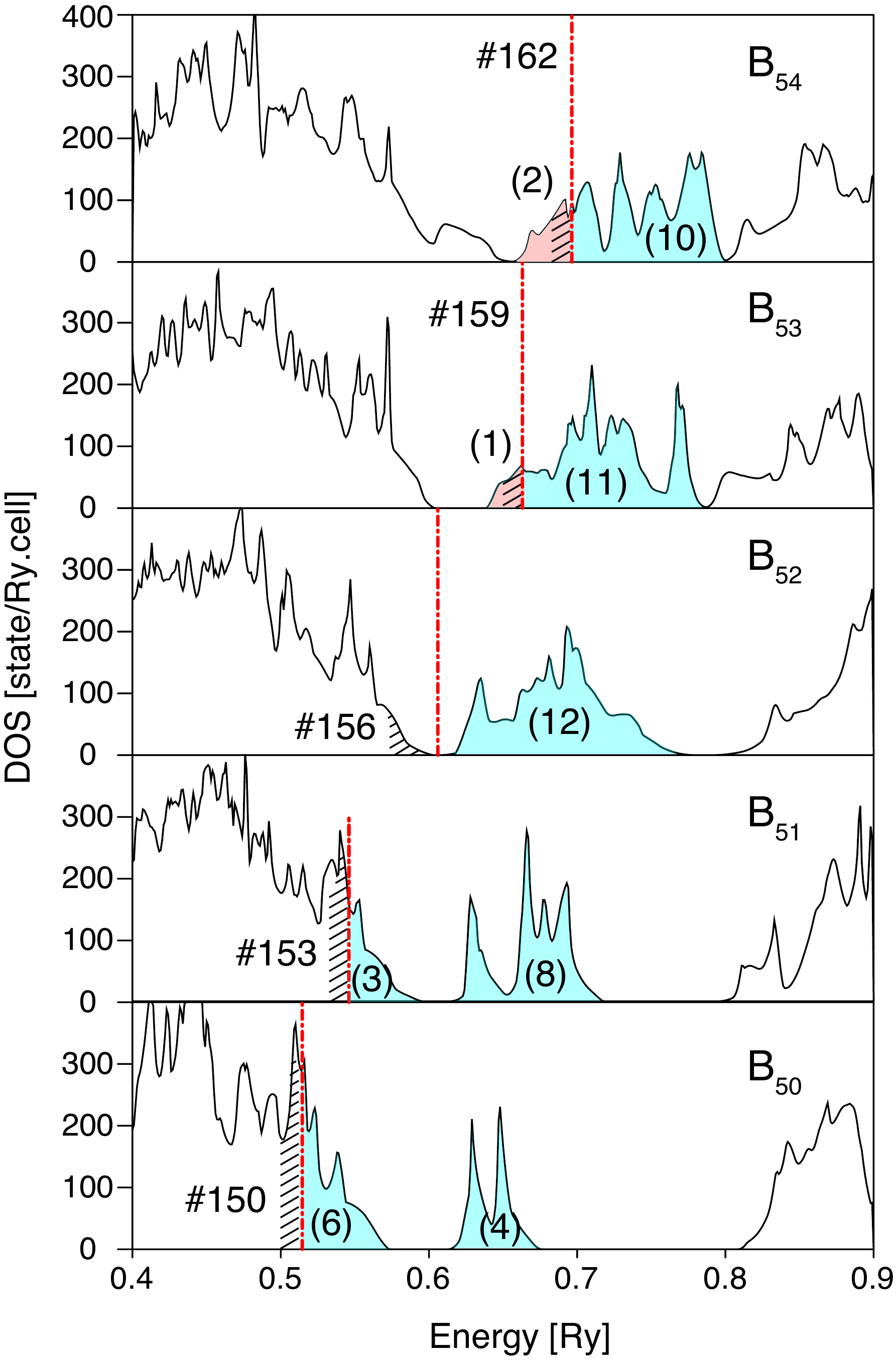}\\
\includegraphics[width=8 cm]{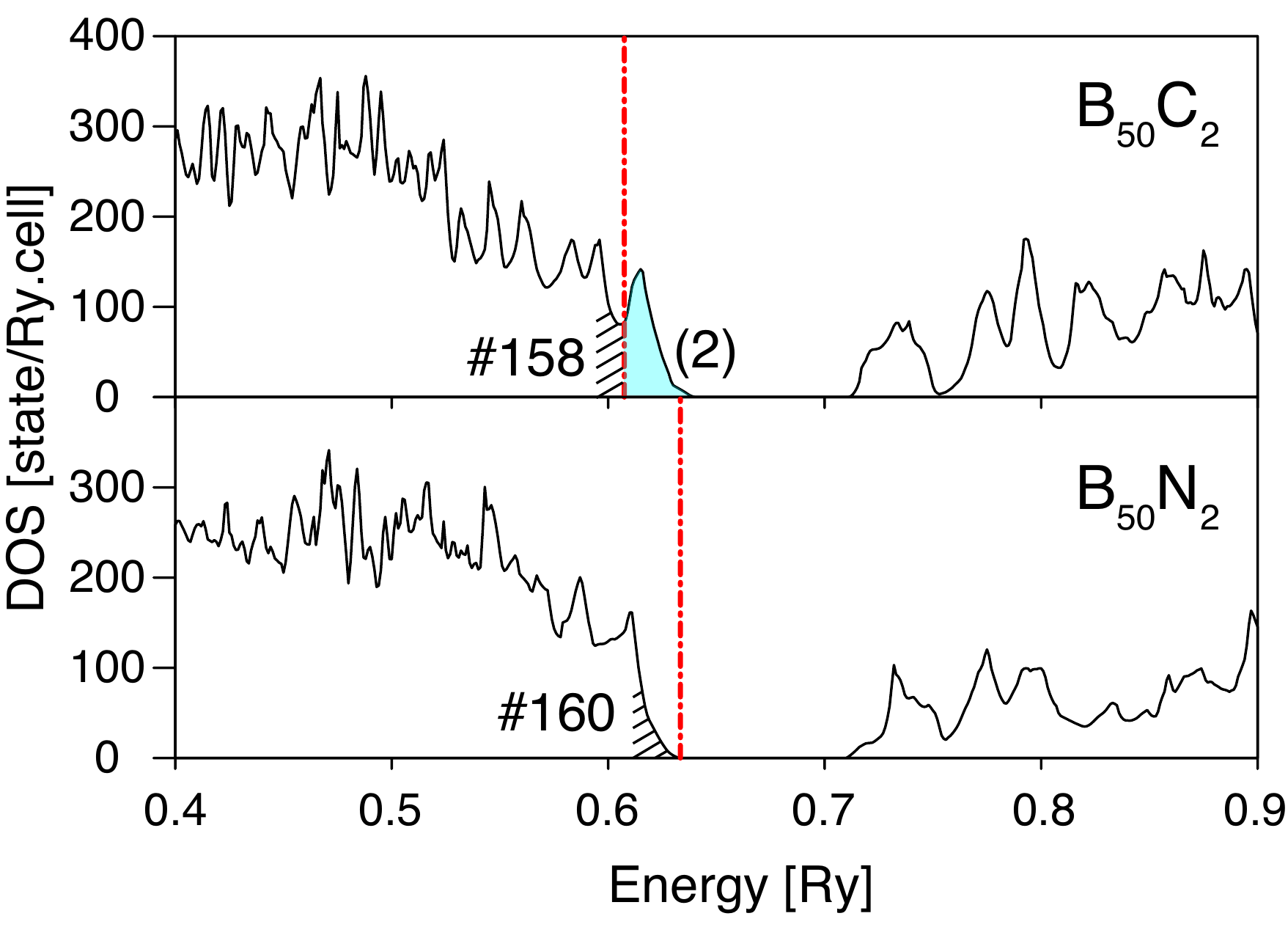}
\caption{(Upper panel) Valence filling for B$_{50+n}$ as the number $n$ of $4c$-site B atoms increases. Only B$_{52}$ can completely fill up the valence bands, but 12 in-gap states remain. This indicates that the conditions for stable covalent bonding are only partially fulfilled. The numbers of occupied states are indicated by the \# symbol. The Fermi levels are indicated by red vertical lines. Light-blue areas indicate unoccupied states, and the numbers of unoccupied states is given in brackets. 
(Lower panel) Similar plot for impurity-containing ${\rm B_{50}C_{2}}$ and ${\rm B_{50}N_{2}}$. ${\rm B_{50}N_{2}}$ can properly fulfill the the conditions for stable covalent bonding, i.e. all valence bands are occupied and no in-gap states are created.} \label{fig:POSfilling}
\end{figure}

The fact that the formation energy is minimized at $n=2$ can be understood by examining the valence filling for different values of $n$.
Figure~\ref{fig:POSfilling} shows the evolution of the density of states (DOS) by successively placing B atoms at $4c$-sites.
For B$_{50}$, there are 160 valence states  (80 bands), and the top 10 states are unoccupied. 
Among these 10 unoccupied states, the lower 6 states are the tail part of the valence band and they are mainly contribution derived from the {\it inter}icosahedral $t-t$ bonds (in the notations of Ref.~\onlinecite{Shirai12}).
The upper 4 states are gap states, which are mainly derived from $p_{z}$ orbitals of two $4c$-site atoms.

As $n$ increases by one, the unoccupied states are filled by the additional 3 valence electrons, almost like a rigid-band shift.
Although it is by no means a rigid-band shift, the valence-electron counting explains this rigid-band-like behavior very well.\cite{Shirai12} In $\alpha$-T boron, each equatorial ($e$ site) B atom of the B$_{12}$ icosahedron  has one {\it inter}icosahedral bond. On placing a B atom at a $4c$ site, four {\it inter}icosahedral $e-e$ bonds are replaced by four tetrahedral bonds of $4c$-site atoms. In total, the number of orbitals is not changed, and consequently three electrons of a B atom are used for filling the unoccupied valence states. 
This way proceeds until $n$=2, where the valence states are completely filled.
Further adding of interstitial atoms creates additional gap states, and it is therefore undesirable. B$_{52}$ with two $4c$-site B atoms is the best configuration to meet the valence requirement in a B$_{50+n}$ series.

This valence filling scheme becomes clearer, when examining the DOS of ${\rm B_{50}C_{2}}$ and ${\rm B_{50}N_{2}}$ (see the lower panel of Fig.~\ref{fig:POSfilling}). For ${\rm B_{50}N_{2}}$, 10 valence electrons from ${\rm N_{2}}$ completely fill the unoccupied states of ${\rm B_{50}}$. Therefore, ${\rm B_{50}N_{2}}$ is the most desirable compound for $\alpha$-T type structures. ${\rm B_{50}C_{2}}$ may be the next desirable form. In contrast, in pure $\alpha$-T boron the unoccupied states are successively filled with increasing $n$, but this has the side effect of also creating gap states.  For the interstitial configurations examined here, we found no configuration that completely separates the valence and conduction bands by using interstitial B atoms only. Probably, there is really no pure configuration without gap states. The significance of this behavior is discussed in Sec.~\ref{sec:frustration}.

Let us now discuss the energy gap of the B$_{52}$ structure. 
Hayami and Otani claimed that B$_{52}$ is a narrow energy-gap semiconductor with an energy gap of 0.07 eV.
(according to a private communication with the authors  Fig.~8 of their paper \cite{Hayami10}, which is expected to indicate this gap, is incorrect.). The corresponding ``gap" is seen in Fig.~\ref{fig:POSfilling}. However, in our opinion the empty bands, indicated by the light-blue area, should rather be regarded as gap states. This means that the LDA band gap is actually 2.7 eV. In experiment, the energy gap was reported to be 1.55 eV by electrical resistivity measurement\cite{Shaw57} and similar values even at high pressures \cite{Ekimov13a}. Unfortunately, even the most comprehensible data handbook\cite{LB-III41c} contains only very little data of the optical properties of $\alpha$-T boron. An optical measurement of a material with a similar structure ${\rm B_{48}Al_{3}C_{2}}$ has a fundamental gap of about 2.0 eV with a tail extending down to 0.5 eV.\cite{Werheit00a} 
According to our experience with boron carbide B$_{13}$C$_{2}$, the fundamental gap itself is large but there are many gap states, which renders the tail part of the conduction band extending deeply into the gap.\cite{Shirai14}
Therefore, it is reasonable to expect a similar gap structure for pure $\alpha$-T boron.

\subsection{Deviation from stoichiometry}
\label{sec:deviation}
The conclusion  that pure $\alpha$-T boron has the stoichiometric composition B$_{52}$, is only provisional. It was obtained by calculations  using  primitive unit cells and will change if we extend the cell size and use supercells. In this section, we examine this possibility.

\subsubsection{Supercell calculations}
\label{sec:supercell}
For studying non-stoichiometry, Ekimov's $\alpha$-T boron is appealing, because only in this case the site occupancy and the composition were determined. 
An interesting point of Ekimov's $\alpha$-T boron is the occupation of IS, as indicated in Table \ref{tbl:alpha-tetra}. The occupancy 0.31 for the $4c$ site amounts to approximately $1 \frac{1}{4}$ atoms per unit cell and 0.06 for the $4g$ site to approximately 1/4 atoms per unit cell. These values suggest an approximate structural model for Ekimov's $\alpha$-T boron as
\begin{equation}
4 \times {\rm B}_{50} + 5 \times {\rm B}_{4c} + {\rm B}_{4g} = {\rm B}_{206},
\label{eq:EkimovModel}
\end{equation} 
which can be realized in a $2 \times 2 \times 1$ supercell. A similar model structure for ${\rm B_{13}C_{2}}$ was studied before.\cite{Shirai14}

\begin{figure}[htbp]
\centering
\includegraphics[width=10 cm]{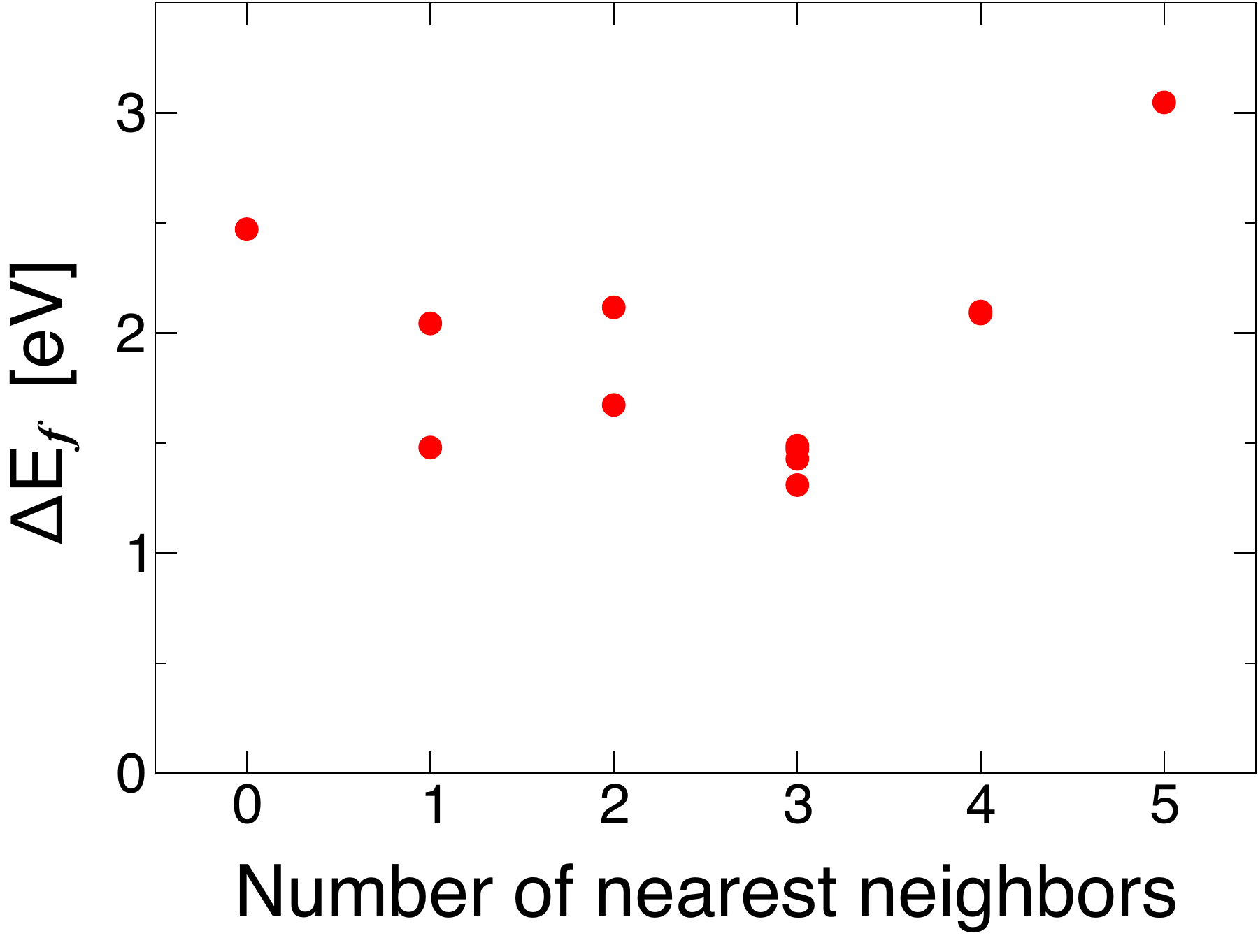}
\caption{The formation energies of different $2 \times 2 \times 1$ supercell  model structures (B$_{206}$) for Ekimov's $\alpha$-tetragonal boron. Overall one finds small positive energy differences, indicating that Ekimov's $\alpha$-tetragonal boron is not the lowest-energy structure. The abscissa is the number of nearest neighbor $4c$-site atoms surrounding a $4g$-site atom and the energy reference state is  $ 4 \times {\rm B}_{52} = {\rm B}_{208}$.} \label{fig:ekimov-confg}
\end{figure}

In this supercell model (\ref{eq:EkimovModel}), one $4g$-site B atom is introduced, and four $4c$-site B atoms are placed, one in each primitive cell. The fifth B atom is randomly placed in one of the four cells, completing the B$_{206}$ model, which corresponds to a deviation from stoichiometry of $\delta=-0.5$. The corresponding formation energy $\Delta E_{f}$, plotted as a function of the number of nearest neighbor $4c$-site atoms surrounding the $4g$-site atom, is shown in Fig.~\ref{fig:ekimov-confg}. In the figure, the energy reference state is B$_{208}$, which corresponds to the stoichiometric B$_{50} + $2B$_{4c}$. 

The formation energy $\Delta E_{f}$ has slight dependence on how many $4c$-site atoms gather around the $4g$ site, but it is not significant.
The energy differences between the B$_{206}$ models and the stoichiometric B$_{208}$ are at least 1.3 eV/atom. The positive $\Delta E_{f}$ shows that the composition of Ekimov's $\alpha$-T is not the lowest-energy composition. Entropic effect cannot change this situation. There are four $4c$ sites in a primitive unit cell. The configurational entropy is maximized at half occupation, $n$=2. And the entropic contribution results only in an improved stability of B$_{52}$, which is already the minimum of the total energy.
Our survey for the lowest-energy structure is by no means exhaustive. But, it is unlikely to find lowest-energy configurations than the present ones by further extending the size of the supercell, when retaining the present structural model. 

\begin{table}
  \caption{The formation energy of interstitial sites and the change of the specific volume $\Delta V$ in a $2 \times 2 \times 1$  supercell model of $\alpha$-T boron. Also in supercell calculations B$_{52}$ is the lowest energy structure. The reference configuration is $ 4 \times ({\rm B}_{50}+ 2 {\rm B}_{4c}) = {\rm B}_{208}$. The multiplicity $g$ of the interstitial sites refers to a primitive unit cell. }
  \label{tbl:HfofB200}
  \begin{tabular}{lcc | crr}
    \hline \hline
    $N_{\rm at}$ ($\delta$) &  & Config. & $g$ & $\Delta E_{f}$ (eV) & \ $\Delta V$(eV/GPa) \\
    \hline
207	&  & $4c$ & 4 & 0.84 & 0.010 \\
  \multicolumn{1}{r}{($-0.25$)} & & $8h$ & 8	& 0.68 & 0.005 \\
	& & $8i$  & 8 & 0.85 & 0.009 \\ \hline
208	&  & $2\times 4c$ & 2 & 0 \\
  \multicolumn{1}{r}{($0$)} &  & $2\times 4c'$ & 2 & 0.20 & -0.022 \\
	& & $4c+8i$ & 4$\times$2 & 0.07 & -0.031 \\
	& & $4c+8h$ & 4$\times$2 & -0.06 & -0.008 \\
	& & $8i+8h$ & 8$\times$2 & 0.22 & -0.043 \\ \hline
209  &  & $3\times 4c$ & 4 & 1.10 & -0.046 \\
 \multicolumn{1}{r}{($+0.25$)} &  & $2\times 4c+8h$ & 4$\times$4 & 0.50 & -0.027 \\
	& & $2\times 4c+8i$ & 4$\times$4 & 1.04 & -0.046 \\
	& & $4c+2\times 8i$ & 4$\times$4 & 0.76 & 0.002 \\
	& & $2\times 4c+8i'$ & 4$\times$4 & 1.15 & -0.052 \\
	& & $4c+8h+8i$ & 4$\times$6$\times$2 & 0.77 & -0.051 \\
    \hline \hline
  \end{tabular}
\end{table}

Next, let us remove some restrictions of the composition and take other IS into account.
A slightly different way of constructing the $2 \times 2 \times 1$ supercell will be employed, now. It is constructed by gathering four unit cells of the lowest-energy configuration B$_{50} + $ 2B$_{4c}$. Then, various combinations of IS are studied in one of the four cells only, and the remaining cells are left to be in the B$_{50} + $2B$_{4c}$ configuration. 
The formation energies of those configurations are listed in Table \ref{tbl:HfofB200}. The energy zero is taken to be that of the stoichiometric B$_{208} = {\rm 4 \times (B_{50} + 2B_{4c})}$. The multiplicity $g_{i}$ of atom configurations for a specific type $i$ of IS is the number of configurations which are energetically degenerate. For convenience, $g_{i}$ is counted within a primitive unit cell. High-energy configurations were omitted beforehand.

Although a low-energy state different from ${\rm B_{50} + 2B_{4c}}$ was found for a combination of $4c$ and $8h$, we still continue to use B$_{50} + 2B_{4c}$ as energy zero. The energy decrease is very small, 0.06 eV. Rather, we should regard $2 \times 4c$, together with $4c+8i$ and $4c+8h$ to form a degenerate ground state of B$_{52}$. In this sense, pure $\alpha$-T boron has non-zero residual entropy even in the stoichiometric composition.

The minimum $\Delta E_{f}$ is still found to be B$_{52}$. From this, we have convinced ourselves that the lowest-energy structure of pure $\alpha$-T boron is indeed the stoichiometric B$_{52}$.
However, the statistical distribution in $\Delta E_{f}$ brings a new feature into the structure B$_{52}$. Furthermore, an imbalance of $\Delta E_{f}$ is found between $\delta=-0.25$ and $+0.25$ (see Table \ref{tbl:HfofB200}).
On the average the case $\delta=+0.25$ has lower energy states than $\delta=-0.25$. In particular, the configuration $2 \time 4c+8h$ has the lowest formation energy among the non-stoichiometric compositions and the value 0.5 eV is not large for high-temperature synthesis. More importantly, the case $\delta=+0.25$ contains larger multiplicities than $\delta=-0.25$. This can yield a positive deviation ($\delta>0$) from B$_{52}$ at high temperatures. In the next section, we will discuss the deviation from stoichiometry $\delta$ at finite temperatures and finite pressure, based on the calculated values in Table \ref{tbl:HfofB200}.

\subsubsection{High-pressure high-temperature properties}
\label{sec:high-pressure}

\begin{figure}[htbp]
\centering
\includegraphics[width=10 cm]{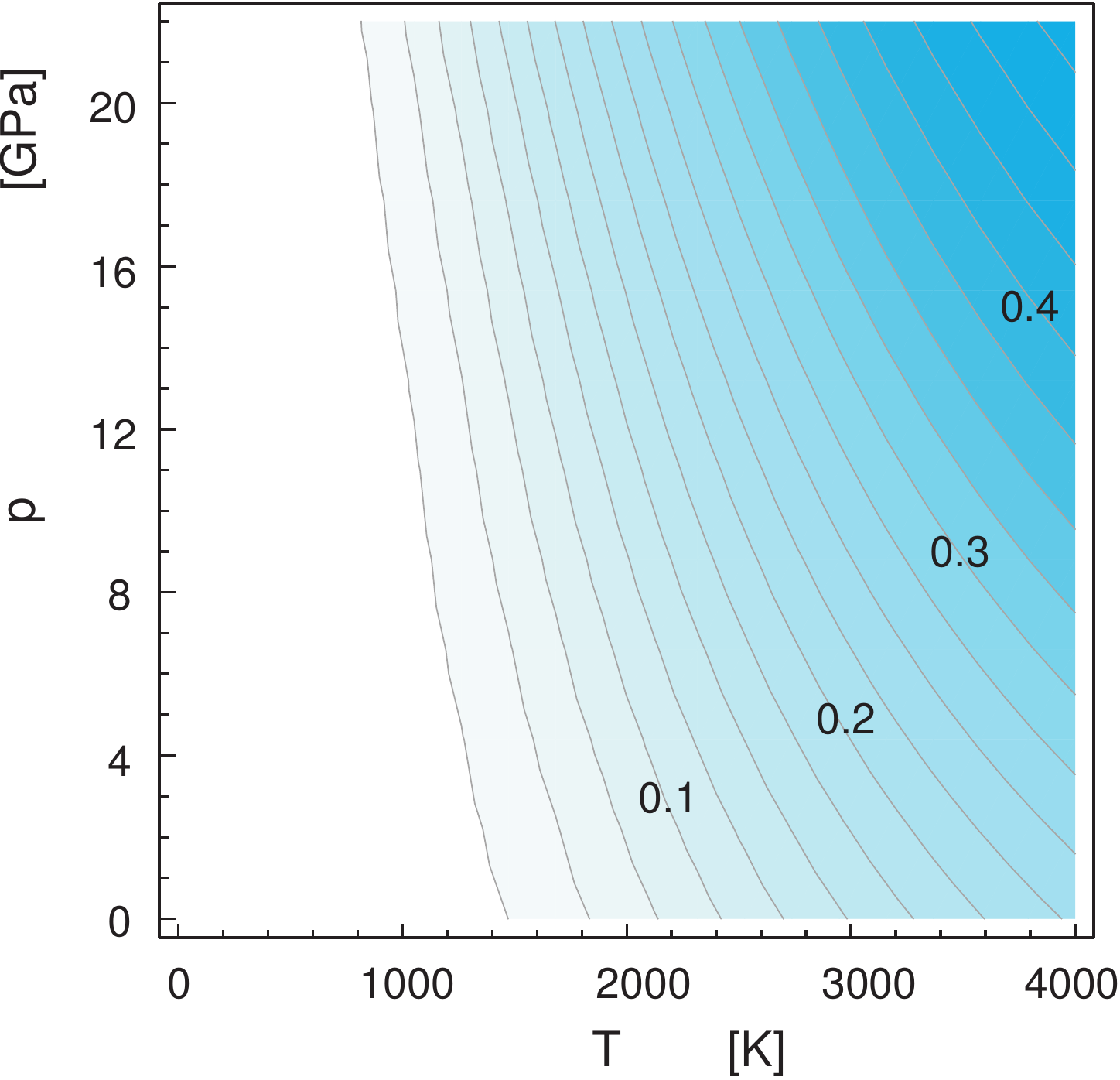}
\caption{Contour plot showing the deviation $\delta$ from the stoichiometric composition B$_{52+\delta}$ as a function of pressure $p$ and temperature $T$. The increment of the contours is 0.025. The results indicate that for high $p$-$T$ synthesis a positive deviation from stoichiometry is to be expected.} \label{fig:delta_pt}
\end{figure}

Considering that most of the recent syntheses of pure $\alpha$-T boron were performed at high pressures $p$, it is important to take the pressure dependence of the formation enthalpy $H$ into account. This can be done by calculating $\Delta H = \Delta E + p \Delta V$.
As shown in Table \ref{tbl:HfofB200}, the magnitude of $\Delta V$ is of the order of 0.02 eV/GPa; so it has a sizable effect on $\Delta H$ at around $p=$ 10 GPa. For the evaluation of $\Delta H$, $\Delta E$ values from supercell calculations were used, because they are more accurate.
Each atomic configuration $j$ in Table \ref{tbl:HfofB200} is characterized by the formation enthalpy $\Delta H_{j}$, the multiplicity $g_{j}$, and the deviation from stoichiometry $\delta_{j}$.
The thermal average of the deviation form stoichiometry $\left< \delta \right> $ is calculated by
\begin{equation}
  \left< \delta \right> = \frac{1}{Z} \sum_{j} \delta_{j} g_{j} \exp \left( -\frac{\Delta H_{j}}{kT} \right),
  \label{eq:AverDelta}
\end{equation}
where $Z$ is the partition function given by,
\footnote{Caution is needed for the values of $\delta_{j}$ when evaluating Eq.~({\ref{eq:AverDelta}}). For example, $\delta_{j}=+1$ should be used for $N_{\rm at}=209$ systems (not 1/4). This is because we are considering the change of the total energy of the crystal that is induced by one interstitial atom; so the size of a supercell is irrelevant.}
\begin{equation}
  Z = \sum_{j} g_{j} \exp \left( -\frac{\Delta H_{j}}{kT} \right).
  \label{eq:PartitionF}
\end{equation}
The quantity $\left< \delta \right> $ has pressure dependence through the enthalply.
It is plotted as function of $p$ and $T$ in Fig.~\ref{fig:delta_pt}.

The obtained values of $\left< \delta \right> $ are of the order of 0.1. With increasing $T$, $\left< \delta \right>$ becomes more positive. At high pressure, negative changes in the specific volume $\Delta V$ are favorable because of LeChatelier's principle.\cite{Yamanaka00,Dekura11} From Table \ref{tbl:HfofB200}, we see that configurations of $N_{\rm at}=209$ are more favorable at high $p$ than configurations of $N_{\rm at}=207$.
Accordingly, $\left< \delta \right>$ increases with $p$.
We see that $\left< \delta \right>=+0.14$ is expected for synthesis conditions of $p=10$ GPa and $T=2000$ K. The corresponding occupancies are 0.20, 0.08, 0.09 for $4c$, $8h$ and $8i$ sites, respectively. Therefore, even at high $T$, the $4c$ site predominates over other IS in pure $\alpha$-T boron.

The predicted value of $\left< \delta \right> \sim 0.1$ is not as large as  $\delta=+1.5$ for $\beta$-R boron, where the deviation is an intrinsic property. This is a consequence of the fact that $\alpha$-T boron has an even number of valence electrons, and thus the driving force for the deviation is weak.
Values of $\left< \delta \right> \sim 0.1$ are also found in ${\rm B_{50}C_{2}}$ and ${\rm B_{50}N_{2}}$ (see Table \ref{tbl:alpha-tetra}). According to the discussion in Sec.~\ref{sec:valenc-filling}, the structures of ${\rm B_{50}C_{2}}$ and ${\rm B_{50}N_{2}}$ meet the valence requirement most properly, and hence they are almost likely to be stoichiometric. Therefore, it is reasonable to find similar values of $\left< \delta \right> $ for pure and C/N-containing $\alpha$-T boron.

At present, there is little experimental information available about $\delta$. Among them, Ekimov's samples are best characterized. They show a negative value $\delta=-0.5$, contrary to the present prediction. By considering their preparation conditions ($p=10$ GPa, and $T=1600^{\circ}$C) this discrepancy is well beyond the error of our calculations. 
As noted before, their samples were synthesized by pyrolysis of decaborane ${\rm B_{10}H_{14}}$. So there is the possibility of hydrogen inclusion, but the authors reported not to detect hydrogen. 
Hyodo's samples with a positive value $\delta=+0.2$ are indeed within the present prediction, however no occupation of the $4c$ site was reported. In this case, the question is whether bulk samples and nanostructures can be compared on the same ground. More reliable measurements on the chemical compositions of the recently discovered $\alpha$-T borons are indispensable for further developments.

\subsubsection{Lattice parameters}
\label{sec:lattice-para}
As indicated in Table \ref{tbl:alpha-tetra}, some trends in the lattice parameters exist between traditional and recently discovered $\alpha$-T borons. When compared with the early-days crystals by Hoard, it is seen that the lattice parameters $a_{0}$ are small for the traditional forms and large for the recently discovered ones. 
In general, the error of GGA for lattice parameters is less than 1 \%, and hence GGA calculations can resolve subtle differences in the lattice parameters which are expected to exist among different $\alpha$-T forms.

\begin{figure}[htbp]
\centering
\includegraphics[width=7.6 cm]{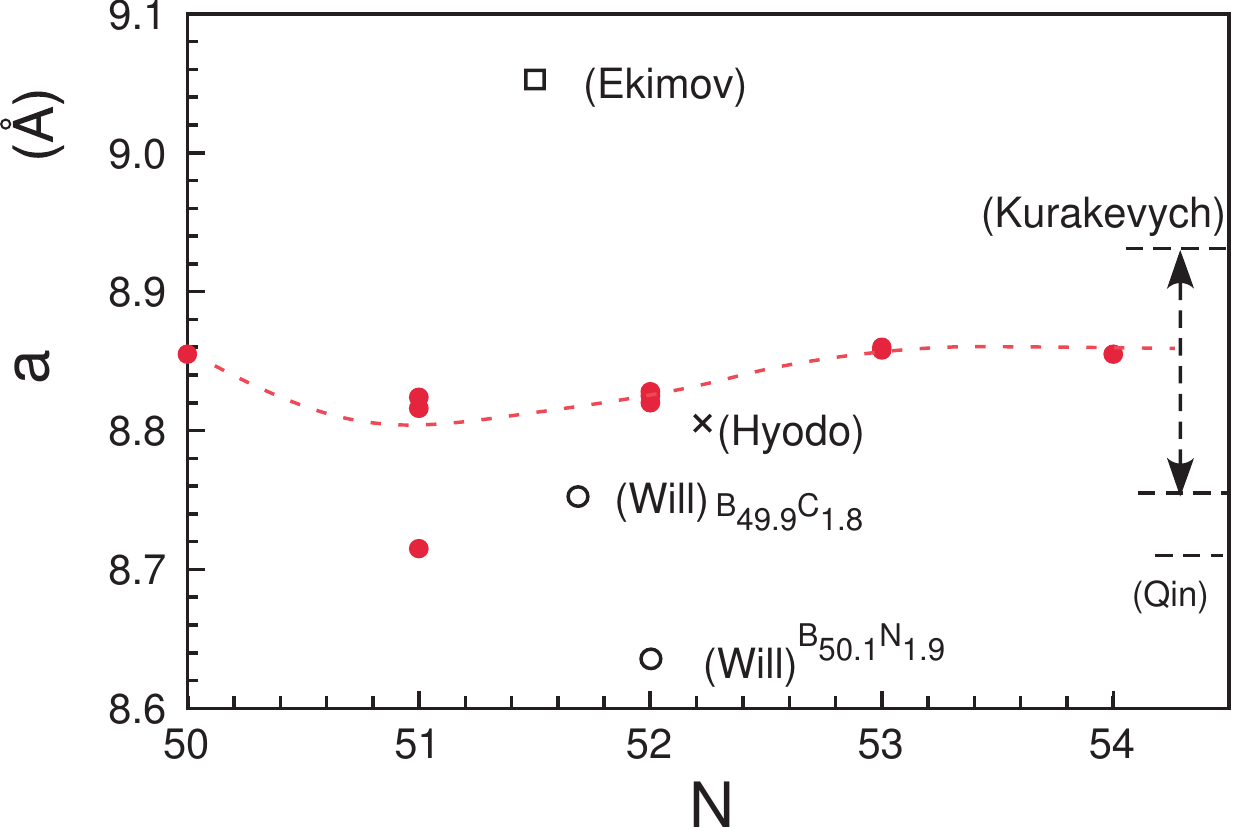}
\hspace{0.4cm}
\includegraphics[width=7.6 cm]{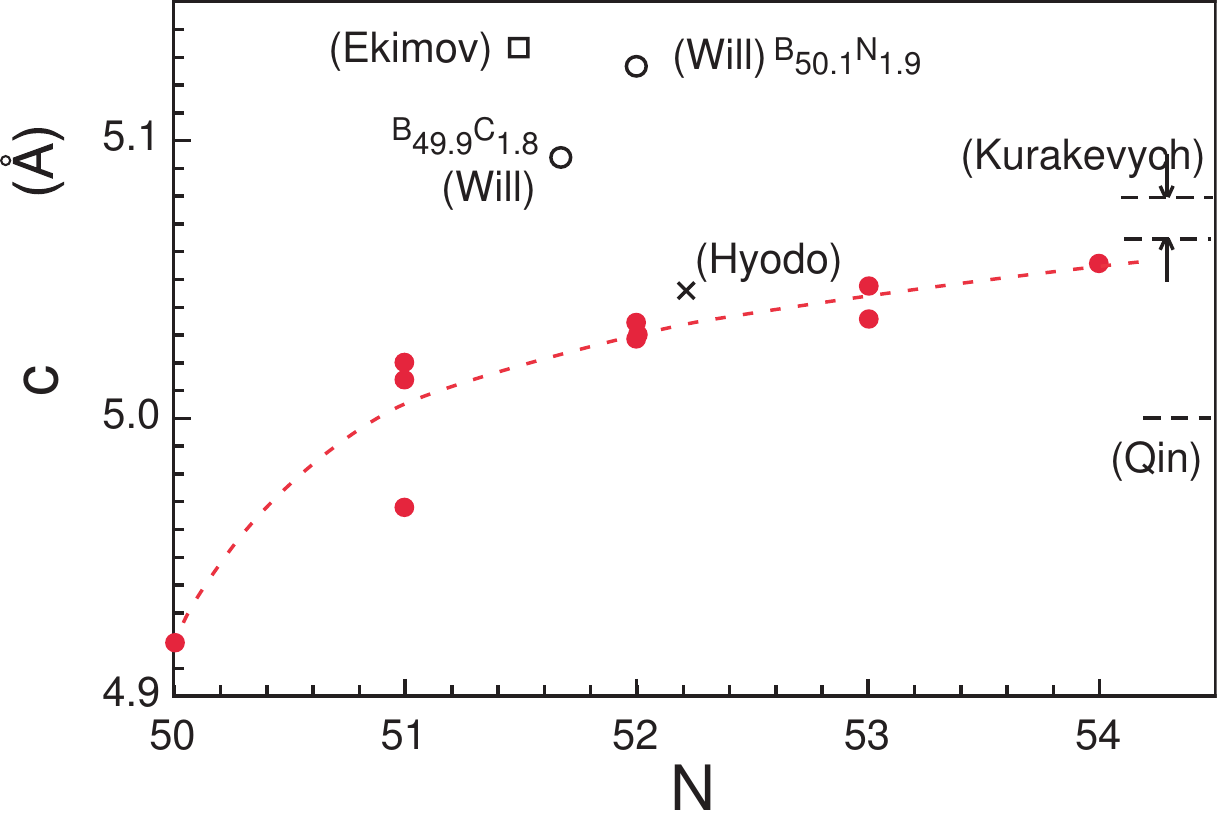}
\caption{The lattice parameters of $\alpha$-T boron as a function of the composition $N$. Red circles indicate the present calculations, symbols with authors's name are experimental data. For Kurakevych's and Qin's $\alpha$-T boron, the chemical compositions are not known and therefore the data are plotted at the rightmost side. Clear differences are discernible between the C/N containing forms and Ekimovs's  $\alpha$-T boron on the one hand and the recently discovered and the calculated forms on the other hand.} \label{fig:lattice}
\end{figure}

Our calculations of the lattice parameters are shown in Fig.~\ref{fig:lattice} as a function of the composition B$_{50+n}$.
The calculated values of $a_{0}$ of pure $\alpha$-T boron are in most cases about 1\% larger than that of the traditional $\alpha$-T boron over a range $n$ = 0 to 4.
When compared with the calculated structures B$_{50}$C$_{2}$ and B$_{50}$N$_{2}$ (not shown in the figure), $a_{0}$ of pure B$_{52}$ is larger than that of B$_{50}$C$_{2}$ by 1.9\% and of B$_{50}$N$_{2}$ by 2.8\%. 
This result is to be expected, because, as discussed in Sec.~\ref{sec:bondnature}, these impurities enhance the covalency of the crystal, resulting in contraction of the cell. 

An opposite trend is found for $c_{0}$; both the calculated values for pure $\alpha$-T boron and the experimental values for the recently discovered ones are smaller than that of the traditional ones. This opposite behavior between $a_{0}$ and $c_{0}$ is a consequence of the distortion of the tetrahedral bonds of the IS. Remember that there is a strong anisotropy between $a_{0}$ and $c_{0}$, despite the tetrahedral arrangement of four icosahedra B$_{12}$. As mentioned repeatedly, the role of interstitial atoms is not small. The bonding of interstitial atoms indeed has to be strong in order to be able to support the local arrangement of icosahedra in the unit cell. An interstitial-site atom, in particular the $4c$ site, has a tetrahedral bonding environment (see Fig.~\ref{fig:POS}). However, these tetrahedral geometries are strongly flatten in the $ab$ plane. Accordingly, the four icosahedra which are connected to the interstitial atom are also deformed in the $ab$ plane. By replacing the interstitial B with 
C or N atoms, the covalency of the related tetrahedral bondings becomes more ideal, {\it i.e.}, isotropic, and the anisotropy between $a_{0}$ and $c_{0}$ is reduced.

Let us check the experimental data. Kurakevych's $\alpha$-T boron contain various concentrations of C including 0 at.\%. They showed an increase in $a_{0}$ as the C content is decreased \cite{Kurakevych12}, which is consistent with our results. Total agreement in $a_{0}$ and $c_{0}$ with our calculation suggests that Kurakevych's $\alpha$-T boron with no C content is pure $\alpha$-T boron. For Qin's case a judgment is more difficult. Although they claimed in their paper \cite{Qin12} that there is a good agreement in the lattice parameters with previous experiments, TEM measurements, which they employed, are generally less accurate. 

The lattice parameter $a_{0}$ of Ekimov's $\alpha$-T boron is evidently larger than the traditional ones. However, his value is too large even when compared with the calculated one for pure $\alpha$-T boron. Judging from the lattice parameters and the negative deviation from stoichiometry $\delta=-0.5$, it is safe to identify Ekimov's $\alpha$-T boron as non-pure boron. Further studies on Ekimov's $\alpha$-T boron are in progress.

\subsection{Geometrical frustration}
\label{sec:frustration}

As discussed in Sec.~\ref{sec:valenc-filling}, and different from impurity-containing forms, a profound feature of pure $\alpha$-T boron is the presence of gap states. Let us consider the electronic nature of the gap states.
For B$_{50}$N$_{2}$, where two $2b$ sites are occupied by two N atoms and half of the $4c$ sites are occupied by two B atoms, the 10 unoccupied states of B$_{50}$ are completely filled by ten valence electrons from two N atoms. On the other hand, for pure $\alpha$-T boron, B$_{50}\ + 2$B$_{4c}$, the lower 6 valence states are filled, leaving 4 topmost states unoccupied. These topmost unoccupied states are filled by the extra 4 electrons in the case of B$_{50}$N$_{2}$, but are left as unoccupied gap states in the pure form.

\begin{figure}[htbp]
\centering
\includegraphics[width=10 cm]{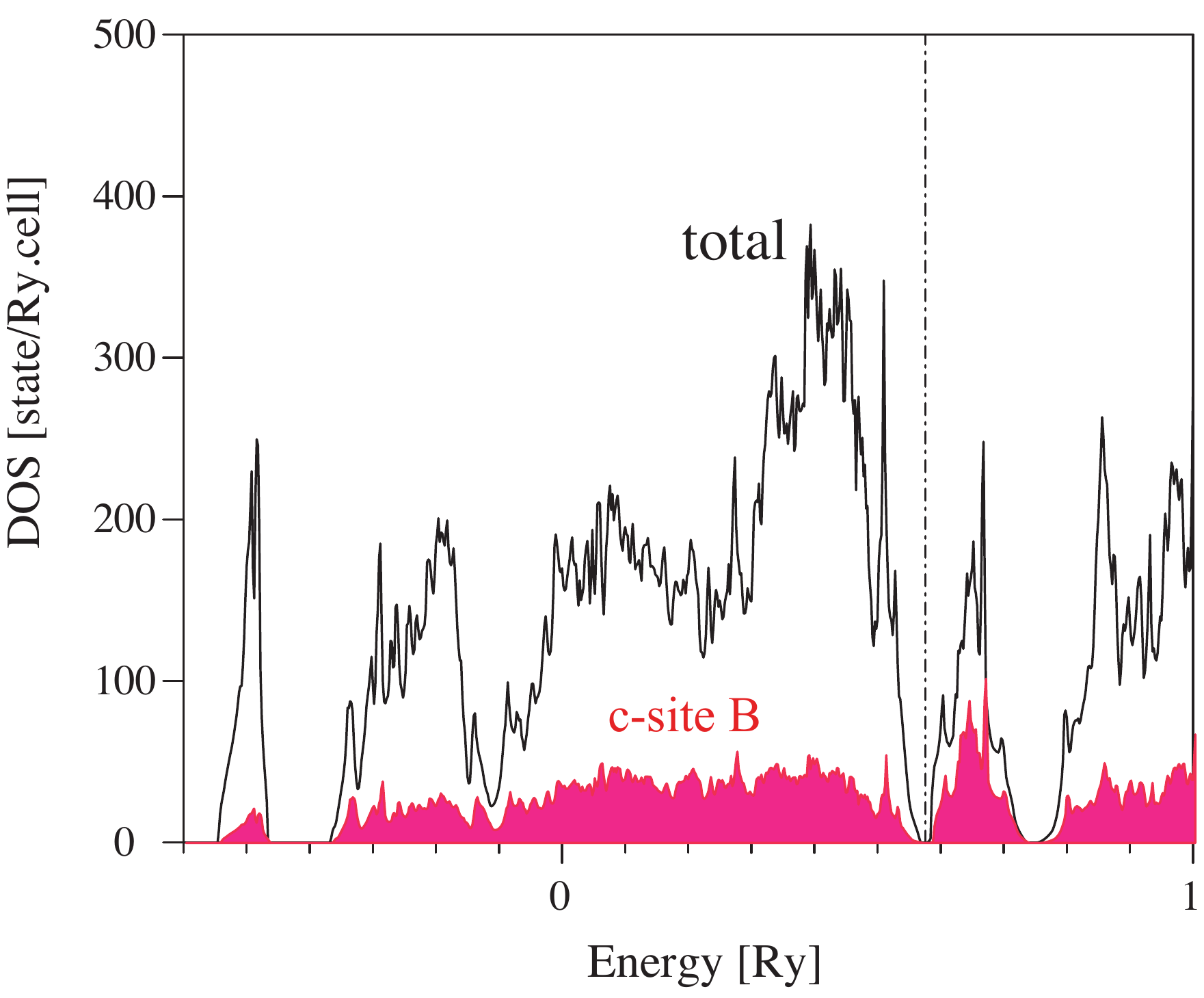}
\caption{Partial density of states (DOS) projected on two $4c$-site B atoms (red area) for B$_{50} + 2$B$_{4c}$ indicates that the in-gap states mainly come from interstitial 4c-site atoms. The black line indicates the total DOS. The scales of partial and total DOS are different.} \label{fig:pDOS2c}
\end{figure}

The electronic nature of these topmost valence states for the pure form B$_{50}+ 2$B$_{4c}$ can be seen by calculating the partial DOS. Figure \ref{fig:pDOS2c} shows the partial DOS with respect to two $4c$-site B atoms. The gap states are mainly coming from the $4c$-site atoms. As mentioned before, the bonding arrangement of the $4c$ site is a strongly flattened tetrahedron; roughly speaking, almost similar to $sp^{2}$ in-plane bonding. The wave functions of the gap states have non-bonding $p_{z}$-orbital character along the direction of the $c$-axis (see Supplemental Material). For the B$_{50}$N$_{2}$ case, the four extra electrons, coming from the two N atoms, can fill these gap states in a similar way as a rigid-band shift.

\begin{figure}[htbp]
\centering
\includegraphics[bb={0 0 2999 2249},width=12 cm]{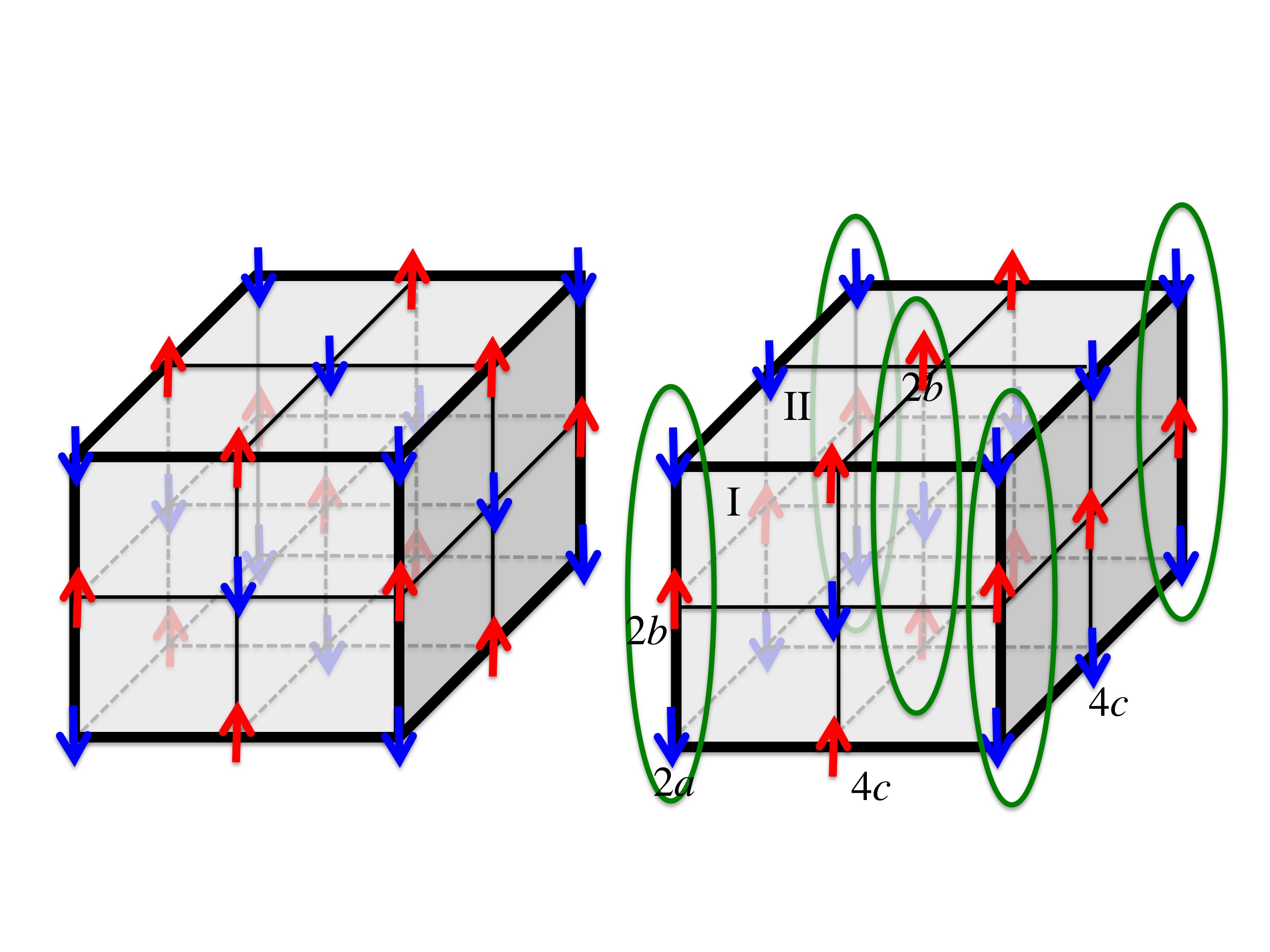}
\caption{The principle of geometrical frustration and its realization in $\alpha$-T boron. (left) A perfectly ordered antiferromagnetic configuration on a cubic lattice. Spin up represents an occupied site, spin down an unoccupied site.  (right) Frustration occurring in B$_{50}+ 2$B$_{4c}$. The illustration shows one primitive unit cell and the vertices of the eight cuboids are the locations of the $2a$, $2b$ and $4c$ sites (compare with Fig.~\ref{fig:struct}). The $2a$ (unoccupied) and $2b$ (occupied) sites, enclosed in green, are fixed in their occupations. The remaining $4c$ sites cannot be arranged in an ordered way. Therefore the system is geometrically frustrated. Roman numbers stand for the index of layers from front to back.}
\label{fig:frustration}
\end{figure}

However, for pure $\alpha$-T boron, the valence filling does not work in this way. When successively adding B atoms to B$_{50}$, each time three unoccupied states are eliminated, until $n=2$. However, this valence filling by B atoms creates new gap states (see Fig.~\ref{fig:POSfilling}).
Further increase of $n$ is even worse as it does not eliminate gap states. In this way, we see that satisfying one bond is connected to the creation of new non-bonding states elsewhere. There is no way to completely remove unoccupied states. This is just the principle of geometrical frustration, which was also found in a similar boron material, {\it i.e.}  $\beta$-R boron.\cite{Ogitsu09}

In a general sense, the icosahedral geometry is intimately related to geometrical frustration.\cite{Sadoc06} But, the special feature of boron is that the geometrical frustration is caused by POS. 
The driving force is strong, because of the covalent nature of the bonding. A pictorial interpretation of the above electronic situation can be obtained by using an idea given by Ogitsu {\it et al} for $\beta$-R boron.\cite{Ogitsu09,Ogitsu10,Ogitsu13} In analogy with antiferromagnetism (AF), it is useful to understand what happens in B$_{50}+ 2$B$_{4c}$.
As noted previously, in boron crystals there is a tendency of interstitial atoms to spatially avoid each other. This situation can be modeled by spin systems. Let us denote an occupied site with spin up, and an empty site with spin down. A perfectly ordered AF phase in a cubic lattice is shown in the left-hand side of Fig.~\ref{fig:frustration}. Let us apply this spin model to B$_{50}+ 2$B$_{4c}$. The right-hand side of the figure corresponds to the adapted spin model. It shows a primitive unit cell as in Fig.~\ref{fig:struct} and the vertices of the eight cuboids are the locations of the $2a$, $2b$ and $4c$ sites. The spin arrangement enclosed in green indicates that the involved spin directions are fixed, that is to say, the $2b$ site is always occupied and the $2a$ site is always unoccupied. Then, we are free to choose the spin directions at the $4c$ sites. Let us consider a spin arrangement in the first layer (denoted as I). The arrangement, as shown in the figure, meets the AF configuration. Every pair 
of neighboring spins is antiparallel. In the next layer (II), the $2a$ and $2b$ sites interchange their positions. Then, an AF spin arrangement within the second layer can be realized as shown in the figure. However, this spin arrangement causes parallel spins between layers I and II. In this way, we will find that there is no perfect ordered state for the occupation of the two $4c$-sites. 

From the geometrical point of view, this frustration can be regarded as arising from the incompatibility of the symmetry, that is, a conflict between the presence of inversion symmetry for icosahedra and the absence of it for the tetrahedral arrangement of four icosahedra in a unit cell. This conflict yields an asymmetry in the occupation between fully occupied $2b$ sites and vacant $2a$ sites, which surround each icosahedron. Two symmetry equivalent $2b$ sites are connected by inversion with respect to the center of an icosahedron, and the same holds for $2a$ sites. This symmetric occupation of $2b$ and $2a$ sites is incompatible with the tetrahedral arrangement of the four icosahedra.

In the stoichiometric B$_{50}+ 2$B$_{4c}$ structure only half of the equivalent $4c$ sites are occupied.  Therefore it has non-zero entropy $S_{0}$ at $T=0$. This residual entropy is (1/52) $\ln 6$ = 0.034 per atom (in units of $k_{\rm B}$, Boltzmann's constant), provided that there is no interaction between $4c$-site atoms. This value is small compared with a value (1/3)$\ln (3/2)$ = 0.135 of water \cite{Ben-Naim09}, which is a prototypical for geometrical frustration, but it is yet a macroscopic amount.
In fact, this simple estimation shows that the residual entropy of $\alpha$-T boron is as large as that of $\beta$-R boron, which was recognized as a geometrically frustrated system. For $\beta$-R boron, the primary IS is a replacement of one atom at the B(13) site with one interstitial atom at the B(16) site. Both of those sites have six equivalent sites. This simple estimate gives the residual entropy (1/105) $\ln (6\times 6)$ = 0.034 per atom. The numerical agreement with the above estimate for $\alpha$-T boron is of course accidental. A more accurate estimate was done by Ogitsu {\it et al}.\cite{Ogitsu10} They determined a value of $S_{0}$ = 0.04 for $\beta$-R boron. The value is very close to the present estimation. Thus the residual entropy of $\alpha$-T boron is comparable to that of $\beta$-R boron. 

\begin{table}
% \begin{table}[htdp]
\caption{A comparison of the site occupancies and the configurational entropy between $\alpha$-T boron synthesized at $p = 10$ GPa and $T=2000$ K (modeled in this work) and $\beta$-R boron (from literature).\cite{Ogitsu10} Both crystals have a similar amount of configurational entropy, caused by geometrical frustration. $N_{\rm at}$ is presented by separating the stoichiometric composition and $\delta$.}
\begin{center}
\begin{tabular}{c| c| l| c| l}
\hline \hline
 Crystal & $N_{\rm at}$ & \multicolumn{1}{c|}{Site occupancy} & $S$ & Notes \\
  &  & \multicolumn{1}{c|}{\%} & $k_{\rm B}$/atom &  \\ \hline
 $\beta$-R boron \cite{Slack88} & 105+1.7 & 74.5 (B13), 27.2(B16), 8.5 (B17) & 0.137 & Equivalent site numbers   \\
   &  & 6.6 (B18), 6.8 (B19), 3.7 (B20) &  & are all 6, except 12 for B(20) \\
 $\alpha$-T boron & 52 +0.14 & 20 ($4c$), 8 ($8h$), 9 ($8i$) & 0.128 &  \\
   \hline
\end{tabular}
\end{center}
\label{tab:entropy}
\end{table}%

Let us now proceed with a more elaborate estimate of the residual entropy. At high temperatures at which the $\alpha$-T boron crystals are synthesized, many POS appear, non-stoichiometric systems are formed (see Sec.~ \ref{sec:high-pressure}), and the configurational entropy increases. We estimate the configurational entropy $S$ by using the following formula,
\begin{equation}
  S = k_{\rm B} \sum_{j} \left[  x_{j} \ln x_{j} + (1-x_{j}) \ln (1-x_{j})  \right],
  \label{eq:entropy}
\end{equation}
where $x_{j}$ is the occupancy of  the IS $j$. By using this formula, it is implied that all the IS are independent.
Experimental values of occupancies for $\beta$-R boron by Slack {\it et al} \cite{Slack88} and our calculated values for $\alpha$-T boron are compared in Table~\ref{tab:entropy}. From these values, we obtain $S$ = 0.137 for $\beta$-R boron and $S$ = 0.128 for $\alpha$-T boron. So the configurational entropy of $\alpha$-T boron is comparable to that of $\beta$-R boron. On one hand this is an unexpected result, because the deviation from stoichiometry in $\alpha$-T boron is only small. On the other hand it is actually reasonable, because the numbers of equivalent sites (the multiplicity $g$) is bigger in $\alpha$-T boron (mostly 8) than in $\beta$-R boron (mostly 6). Moreover, the number of atoms per unit cell in $\beta$-R boron is twice as large as that in $\alpha$-T boron.

Based on these arguments, $\alpha$-T boron can be considered as a new member of the group geometrical frustrated boron systems, because (i) an unavoidable conflict between the elimination of the existing unoccupied states and the creation of new unoccupied states, and (ii) the existing of a macroscopic amount of residual entropy. An interesting point is that the geometrical frustration of $\alpha$-T boron does not require non-stoichiometry, differently from the situation in $\beta$-R boron. This makes the theoretical analysis simpler. Recently, a novel view, the so-called \textit{correlated disorder} has appeared, which is a universal approach connecting different classes of disorders, that is, frustrated over-constraint and configurational under-constraint.\cite{Keen15} The structure of $\alpha$-T boron is a particularly a good system for developing this view, because the structure seems to have both characters of disorder in it, because there is no way to perfectly satisfy covalent conditions (over-
constraint) and there are many ways of arranging POS (under-constraint).

\section{Conclusions}

We have examined the possibility of the existence of non-stoichiometric, pure $\alpha$-T boron. Supercell calculations have shown that the lowest-energy state of pure $\alpha$-T boron, if it exists, is the stoichiometric B$_{52}$ with two occupied  $4c$ sites. Therefore, a deviation from stoichiometry occurs only by entropic effects at high temperatures.
This is different from $\beta$-R boron, where non-stoichiometry is a property of the lowest-energy state. This difference essentially comes from the fact that B$_{52}$ has an even number of electrons and therefore the driving force for a deviation from stoichiometry is relatively weak.
Finite temperatures cause a deviation from stoichiometry B$_{52+\delta}$ by a small positive amount, with $4c$ site atoms as the main interstitial species. 
For high-pressure and high-temperature synthesis under reported conditions,  $\delta$ is estimated to be around $\delta$=0.1 to 0.2.

The present results on the site occupancies, the $\delta$ values as a function of pressure and temperature, and the lattice parameters provide a good test for the experimental identification of pure $\alpha$-T boron. The dominant interstitial site of the pure form is $4c$, while for C- or N-containing forms the $2a$, $8h$ and $8i$ sites are favored.
Judging from the limited experimental data, we conclude that Ekimov's $\alpha$-T boron is probably not a pure form, because of its negative $\delta$ together with other inconsistencies. Judging from the lattice parameters, other recently discovered $\alpha$-T boron forms could indeed be pure ones, though not all. For a final conclusion, however, more detailed information about the chemical composition and the site occupancies are required.

Despite of B$_{50}$ + 2B$_{4c}$ having a band gap, the valence bond requirement is not ideally fulfilled. Unoccupied $p_{z}$ orbitals at $4c$ sites form gap states that cannot be eliminated by adding further interstitial atoms.  There is no way to fully satisfy the conditions for covalent bonding. Despite of being stoichiometric, B$_{52}$ has a macroscopic amount of residual entropy that is as large as that of $\beta$-R boron. Because of having such a lowest-energy state, macroscopic residual entropy and no atomic arrangement that can satisfy the conditions for covalent bonding, pure $\alpha$-T boron is identified as a geometrically frustrated elemental crystal.

\section*{Acknowledgments}
The authors thank K. Kimura and W. Hayami for initiating our interests in this subject.
We also thank E. Ekimov for discussing the characterization of his samples. 
JK acknowledges financial support from the German Research Foundation (DFG) (project KU 2347/2-2).
HE and JK acknowledge financial support from the German Academic Exchange Service (DAAD) (project 54368630)

% \section*{References}

% \bibliography{boron,added}

\end{document}